\documentclass[aps,pra,twocolumn,groupedaddress,showpacs,showkeys]{revtex4}

\usepackage{graphicx}
\begin{document}

\newcommand{\be}{\begin{equation}}
\newcommand{\ee}{\end{equation}}

\title{Making  precise predictions of the Casimir force between metallic plates via  a  weighted Kramers-Kronig transform}

\author{ Giuseppe Bimonte}
\email[Bimonte@na.infn.it]
\affiliation{Dipartimento di Scienze Fisiche Universit\`{a} di
Napoli Federico II Complesso Universitario MSA, Via Cintia
I-80126 Napoli Italy and INFN Sezione di Napoli, ITALY\\
}

\date{\today}

\begin{abstract}

The possibility of making precise predictions for the Casimir
force is essential for the theoretical interpretation of current
precision experiments on the thermal Casimir effect with metallic
plates, especially for sub-micron separations. For this purpose it
is necessary to estimate very accurately the dielectric function
of a conductor along the imaginary frequency axis. This task is
complicated in the case of ohmic conductors, because optical data
do not usually extend to sufficiently low frequencies to permit an
accurate evaluation of the standard Kramers-Kronig integral used
to compute $\epsilon(i \xi)$. By making important improvements in
the results of a previous paper by the author, it is shown that
this difficulty can be resolved by considering suitable weighted
dispersions relations, which strongly suppress the contribution of
low frequencies.  The weighted dispersion formulae presented in
this paper permit to estimate accurately the dielectric function
of ohmic conductors for imaginary frequencies, on the basis of
optical data extending from the   IR to the UV, with no need
of uncontrolled data extrapolations towards zero frequency that
are instead necessary with standard Kramers-Kronig relations.
Applications  to several
  sets of data for gold films are presented to demonstrate
viability of the new dispersion formulae.
\end{abstract}


\maketitle

\section{INTRODUCTION}

The Casimir effect, and more in general dispersion forces, are the
current object of much theoretical and experimental interest. The
reasons for the continuing interest in this well-established field
are numerous, and are connected with both fundamental and applied
science. For an updated overview of the subject, the reader is
referred to  recent reviews \cite{parse,bordag,Mohid}.

After fifty years of slow progress, the field of Casimir physics
received a strong push in the last decade, as a result of a new
wave of experiments \cite{lamor} which succeeded for the first
time in measuring the tiny Casimir force with unprecedented
precision. The new measurements provided a definitive confirmation
of this phenomenon, but at the same time experiments utilizing
metallic surfaces (which  are used in the majority of current
experiments) have raised unexpected theoretical puzzles,  for the
so-called thermal Casimir force. The problem is whether in the
computation of the Casimir force at finite temperature conduction
electrons in the plates  should be described as a collisionless
plasma, similarly to what is done in infrared optics, or if
relaxation processes should be included
\cite{bordag,sernelius,bimonte2,intravaia}. The two approaches
have been dubbed in the Casimir literature as  plasma and Drude
prescriptions, respectively. The problem is rather subtle indeed,
with non trivial implications in thermodynamics and statistical
physics that are not yet fully understood. In fact, it has been
found that inclusion of relaxation effects in the idealized case
of two plates with no imperfections leads to violation of the
Nernst theorem \cite{bezerra}, while no such violation occurs when
any amount of impurities is present \cite{brevik}. It has been
shown however that neglect of relaxation processes leads to
contradiction with the Bohr-van Leeuwen theorem of classical
statistical physics \cite{bimonte4}. From an experimental point of
view, the resolution of the problem requires observation of the
thermal Casimir force between two metallic plates, because the two
alternative theoretical pictures give different predictions for
its magnitude. The experimental situation is perplexing, and at
the moment of this writing there appears to be a striking
contradiction between the results of different experiments by
different groups. A series of experiments using microtorsional
oscillators by the Purdue group \cite{decca}, providing at this
time  the most precise measurements of the Casimir force between
metallic bodies in the separation range from 160 nm to 750 nm,
were shown to be in agreement with the plasma prescription and to
rule out the Drude prescription. Also a large distance torsion
balance experiment \cite{masuda} in which the Casimir force was
measured in the separation range from 0.48 to 6.5 $\mu$m, obtained
results that are in agreement with the assumption of ideal metal
plates, and are in contradiction with the Drude model.  On the
contrary, a new experiment \cite{lamor3} in which the Casimir
force between a spherical lens and a flat plate was measured in
the range from 700 nm to 7 $\mu$m by a torsional balance,  was
claimed to be fully consistent with the Drude prescription, and to
rule out the plasma one. This state of things calls for a careful
investigation, to see if the contradiction can be resolved.

It is well known by now \cite{bordag} that the theoretical
interpretation of precision Casimir experiments is a difficult
problem, because several possible sources of error must be
considered, that include for example consideration of unavoidable
imperfections of the surfaces, like roughness and other
uncontrolled geometric distortions  \cite{umar}, the possible
presence of residual electrostatic forces etc. We shall not be
concerned with these issues here, and for a thorough discussion of
the problem of theory-experiment comparison in Casimir
experiments, we address the reader to the recent work \cite{geyer,
geyer2}. In this paper we shall focuss our attention on the
influence of the optical properties of the the plates, described
by their dielectric function $\epsilon(\omega)$,   which by far
constitutes the dominant factor affecting the magnitude of the
Casimir force, especially at sub-micron separations.
 The effect
of the optical properties of the plates  on the Casimir force must
be estimated very precisely, if one seeks to discriminate between
alternative prescriptions for the thermal Casimir effect, because
the predicted magnitudes of the force resulting from different
approaches  differ   by just a few   per cent at  separations
below one micron, which is the range investigated in the
experiments by the Purdue group.

According to the theory of dispersion forces \cite{parse}  in
order to compute the Casimir force between two dielectric
(non-magnetic) plates of any shape at temperature $T$ one needs to
know the electric permittivities $\epsilon(i \xi_n)$ of the
materials constituting the plates, evaluated at certain
temperature-dependent discrete imaginary frequencies $\xi_n= 2 \pi
n k_B T/\hbar$, where $n$ is a non-negative integer, known as
Matsubara frequencies.
Getting accurate values for the dielectric function $\epsilon(i
\xi)$ is not easy, in particular when ohmic conductors are
considered. This is so because the imaginary-frequency dielectric
function $\epsilon(i \xi)$ cannot be directly measured by any
experiment, and therefore it has to be computed using
Kramers-Kronig (KK) dispersion relations, starting from optical
data of the involved materials. We shall specifically consider the
case of gold, which is the material  of interest in the recent
experiments probing the thermal effect. Until a few years  ago
\cite{capasso, lisanti,iannuzzi} it was customary   to compute
$\epsilon(i \xi)$ by using the data for gold tabulated in the
handbook \cite{palik}. However, in the most recent experiments the
new practice has been adopted of computing the quantities
$\epsilon(i \xi_n)$ on the basis of optical data measured directly
on the samples used in the Casimir measurements. The necessity of
doing this was first stressed in \cite{lamor2} and it stems from
the fact that  optical properties of  thin polycrystalline gold
films, typically used in Casimir experiments, are  much dependent
on the used deposition method. It has been claimed
\cite{Piro,sveto} that possible sample-to-sample variations of the
optical properties may engender significant variations in the
Casimir force,  possibly larger than 10 per cent at submicron
separations. While  the large differences in the Casimir force
reported in \cite{sveto} probably do not arise for carefully
prepared gold films like those used in the experiments
\cite{decca}, it seems correct to state that the optical
properties of the used samples should be carefully characterized,
if per cent precision in the theory-experiment comparison is aimed
at.

A big practical problem   remains however because, due to the
characteristic $1/\omega$ singularity displayed by  the dielectric
function of ohmic conductors, the standard KK integral used to
compute $\epsilon(i \xi)$ (Eq. (\ref{dispbis}) below) receives a
very large contribution from frequencies below the IR, for which
it is very difficult to collect optical data  for thin metallic
films. The remedy to this problem, initially adopted in
\cite{lambrecht} and followed afterwards by all authors, consists
in extrapolating the available data below the minimum accessible
frequency $\omega_{\rm min}$, in order to estimate the KK integral
in the interval $0 < \omega <\omega_{\rm min}$ where optical data
are not directly available. The extrapolation is usually done by
means of the simple Drude formula for the dielectric function of
ohmic conductors. It is important to observe that this
contribution is usually very large (it easily accounts for more
than fifty per cent of the KK integral) and therefore the values
of the Casimir force that are obtained by this procedure are very
sensitive to experimental uncertainties in the values of the Drude
parameters. For example, in Ref. \cite{sveto} it was estimated
that experimental uncertainties in the Drude parameters may imply
by themselves an uncertainty in the Casimir force as large as 1
$\%$ at 100 nm.

In our  opinion, the reliability of the whole procedure described
above can be questioned, because in reality there is no way to
know how accurately the Drude model   describes the dielectric
function of   gold films, in the wide range of frequencies for
which it is used in Casimir computations.
It seems to us that there is no way to quantify the error
determined by possible deviations of the actual dielectric
function from the Drude model. In our judgment, in order to obtain
fully reliable predictions of the Casimir force with metallic
plates it is necessary to find means of reducing the impact of the
Drude extrapolation.  In principle    the straightforward way to
achieve this goal would be by further extending optical data on the
low frequency side. It is unlikely however that much progress can
be made in this way, since it has been estimated that a
determination of the Casimir force between two gold plates with an
error of half per cent at a separation of 150 nm solely on the
basis of measured optical data, would require extending optical
measurements till wavelengths in the millimeter region, which is a
very difficult  thing to do.

An alternative approach, which does not require   further
extension of optical measurements beyond the limits of  currently
available optical data, was proposed recently by the author in
Ref. \cite{bimonte}. The idea is to introduce in the dispersion
relations used to compute $\epsilon(i \xi)$ suitable analytical
weights $f(z)$, called by us "window" functions, that suppress the
contribution of frequencies for which no optical data are
available. We should observe that consideration of weighted
dispersion relations is well documented in the optical literature,
where they have been exploited for example to derive new important
relations connecting the refraction index to the extinction
coefficient. For details, we address the reader to Chapter 3 of
\cite{palik}.
In our first work \cite{bimonte} we demonstrated that by taking
weights $f(z)$   that vanish both at zero and at infinity, it is
indeed possible to suppress very effectively the contribution of
frequencies outside the data interval $\omega_{\rm min} < \omega <
\omega_{\rm max}$. A simple family of window functions,
parametrized by two integers $p$ and $q$ (with $p<q$) and by a
complex  frequency $w$, was offered there, and it was shown that for
suitable choices of the parameters the error made by truncating
the integral to the frequency range $\omega_{\rm min} < \omega <
\omega_{\rm max}$ can be made negligible at both ends of the
integration domain, in such a way that a precise determination of
$\epsilon(i\xi)$ is possible, without extrapolating the optical
data outside the interval where they are available. The key
feature of the generalized dispersion relation that  makes this
possible is that, differently from the standard KK relation  which
only involves $\epsilon''(\omega)$, the generalized relations
involve both $\epsilon'(\omega)$ and $\epsilon''(\omega)$, and
therefore they exploit the full information delivered by   optical
data. This is not a problem in principle if one considers that
optical techniques exist, like ellipsometry, that permit to
measure independently both $\epsilon'(\omega)$ and
$\epsilon''(\omega)$ with good precision. Such a technique was
used for example in \cite{sveto}.


The analysis in \cite{bimonte} was unrealistic however, in that it
assumed that  optical data in the interval $\omega_{\rm min} <
\omega < \omega_{\rm max}$ were known with an infinite precision.
Real optical data of course are not like that, as they are
affected by experimental errors of various nature. That the
analysis in \cite{bimonte} was oversimplified has been pointed out
recently in \cite{geyer}, where the windowed dispersion relations
were tentatively applied to the tabulated data for gold of Ref.
\cite{palik}. Using the same window functions and the same window
parameters used in \cite{bimonte}, the authors of \cite{geyer}
obtained negative values for the quantity $\epsilon(i\xi)$ in
certain intervals of the imaginary axis, and this is absolutely
unacceptable because on general grounds one knows that along the
imaginary axis the electric permittivity of a causal medium must
be positive \cite{lifs2}. In an attempt at understanding their
findings, the authors of Ref. \cite{geyer} observed that the
handbook \cite{palik} combines data from several distinct
experiments, using gold films prepared by different procedures.
Since, as we said earlier, optical properties of gold films depend
significantly on the deposition method,  the operation of
combining optical data for  films having exceedingly different
properties (that this is indeed the case with the tabulated data
is shown in Sec. V below) may in principle spoil the overall KK
consistency of the data, which is an essential precondition for
dispersion relations to work well. We agree with the authors of
Ref. \cite{geyer}, though, that this potential problem with the
tabulated data of Ref.\cite{palik} cannot explain the obtained
negative values of $\epsilon(i\xi)$. They argue that the main
reason is the possible strong sensitivity of the windowed
dispersion relations  to   experimental errors in the optical
data, in particular in the real part of the permittivity. If an
exceedingly large amplification of small uncertainties in the
optical data were an unavoidable feature of our windowed
dispersion relations, their practical utility would be severely
diminished, of course.  The issue of error propagation is  an
important point deserving a detailed analysis,  and we address it
in the present paper. We have performed a Monte Carlo simulation
to determine how random errors in the optical data  are propagated
by the windowed dispersion relations.  The results of these
simulation confirm the suspicion of the authors of Ref.
\cite{geyer}, and explain their findings, revealing that the
choice of the window  functions  that was made  in \cite{bimonte},
was very unfortunate indeed, because it leads to a huge
amplification of   errors in the optical data. Trying to resolve
the problem, we realized that the key property protecting the
standard KK relation from this sort of instability is the fact
that it involves a positive definite kernel. This property was
badly violated by the weighted dispersions relations considered in
\cite{bimonte}, which are instead characterized by kernels
attaining large negative and positive values. After several
attempts, we could eventually fix this problem and we found a new
class of weight functions, presented in this paper, that lead to
positive kernels when applied to ohmic conductors. In this paper
we demonstrate that the new weighted dispersion formulae do not
suffer from large error propagation, and therefore they can be
used to reliably compute $\epsilon(i \xi)$, and in turn the
Casimir force, solely on the basis of optical data extending from
the   IR to the UV, with no need of further extending optical
measurements to lower frequencies.

The plan of the paper is as follows: in Sec. II we review  the
  KK  formula routinely  used to compute the dielectric function
$\epsilon(i \xi)$, and we discuss its weaknesses in the case of
ohmic conductors. In Section III, we review briefly the weighted
or "windowed" dispersion relations that were introduced in
\cite{bimonte}, and we present a new class of weights or "window
functions" leading to dispersion relations with positive kernels
when applied to conductors. In Sec. IV we present our numerical
computations which demonstrate how the window functions considered
in our first work  suffer from a strong instability, and we show
that the new weights  resolve this problem. In Sec. V we apply the
windowed dispersion relations to tabulated optical data for gold
of Ref. \cite{palik}, and in Sec. VI to some of the data in
\cite{sveto}. In Sec. VII we use the weighed dispersion relations
to derive an alternative expression for the so-called generalized
plasma model that has been recently advocated as providing the
correct description for ohmic conductors in the Casimir effect.
Finally, Sec VIII contains our conclusions and a discussion of the
results.

\section{Dielectric function at imaginary frequencies}

As it is well known from the theory of dispersion forces
\cite{parse} the Casimir interaction between two macroscopic
bodies at temperature $T$ depends on the dielectric functions of
the two bodies, evaluated for a discrete set of imaginary
Matsubara frequencies $\xi_n=2 \pi n k T/\hbar$, where $n$ is any
non-negative integer.  Finding means of accurately computing the
quantities $\epsilon(i \xi_n)$ embodying the material dependence
of the Casimir force, is the primary goal of this work. We shall
see below that this is not a trivial task, in particular when
ohmic conductors are considered. As we said earlier, it is
impossible to directly measure $\epsilon(i \xi)$ because any
conceivable optical experiment can only determine the dielectric
permittivity $\epsilon(\omega)$ along the real frequency axis.
Knowledge of optical data along the real axis can be exploited to
compute $\epsilon(i \xi)$ by different mathematical procedures,
depending on the sought accuracy. For moderate precisions, at five
per cent level or so, the simplest method is to make oscillator
models of the optical data $\epsilon(\omega)$ for the material of
interest. The explicit analyticity of such analytical models then
permits to obtain the quantities $\epsilon(i\xi_n)$ by direct
substitution $\omega \rightarrow i \xi_n$ into the oscillator
formula. However, this straightforward procedure is not accurate
enough   for the purpose of interpreting the most recent precision
Casimir experiments. When higher precision is needed, the
quantities $\epsilon(i \xi_n)$ must be determined directly on the
basis of   optical data, without making recourse to simplified
analytical models for the latter.  This can be done by exploiting
dispersion relations entailed by causality of the electric
permittivity, which   relate the unmeasurable quantities
$\epsilon(i \xi_n)$ to the measurable real-frequency permittivity
$\epsilon(\omega)$. The dispersion relation that has been utilized
until now in the literature is the following KK relation \be
\epsilon( i \xi)-1=\frac{2}{\pi}\int_0^{\infty} d \omega
\frac{\omega\,
\epsilon''(\omega)}{\omega^2+\xi^2}\;,\label{dispbis}\ee which is
valid for all materials, like insulators and ohmic conductors,
whose electric permittivity diverges at most like $1/\omega$ at
zero frequency.

As we explained in the Introduction application of the above
formula to ohmic conductors meets with difficulties. We recall
first that the important range of Matsubara frequencies that need
be considered in Casimir computations  depends on the separation
$a$ between the two bodies. In practice \cite{bordag}, for an
accurate determination of the Casimir force it is sufficient to
consider Matsubara frequencies up to a maximum value of about ten
times the characteristic frequency $\omega_c=2/(2 a)$. For
separations $a$ larger than 50 nm, which are the ones probed by
current experiments, is is therefore sufficient to consider
Matsubara modes with frequencies $\xi$ ranging from $0.16$
eV/$\hbar$, representing the frequency of the first Matsubara mode
at room temperature, up to a maximum frequency of ten eV/$\hbar$
or so. Now we see from Eq. (\ref{dispbis}) that in order to
evaluate $\epsilon(i \xi)$ at any   imaginary frequency $\xi$  it
is in principle necessary to know $\epsilon''(\omega)$ at all
frequencies $\omega$. Unfortunately such a complete knowledge of
$\epsilon''(\omega)$ is never possible, because optical data are
always   restricted to some finite frequency range $\omega_{\rm
min} < \omega < \omega_{\rm max}$, starting from a non-zero
minimum frequency $\omega_{\rm min}>0$. Consider for example the
data for gold collected in the handbook \cite{palik}, which have
been routinely utilized to interpret Casimir experiments using
gold plates, including the recent experiments \cite{decca}. These
data, which we shall describe more accurately later on, cover a
range of frequencies extending from 0.125 eV/$\hbar$,
corresponding to the   IR, till $10^4$ eV/$\hbar$.
Unfortunately, this range is not sufficiently wide to permit an
accurate estimate of the integral on the r.h.s. of Eq.
(\ref{dispbis}) in the relevant range of imaginary frequencies
$\xi$, which as we said extends from 0.16 eV/$\hbar$ to ten
eV/$\hbar$. In fact there is no   difficulty on the high frequency
side, because fall-off properties of $\epsilon''(\omega)$ entail
that for all $\xi$'s in this range, real frequencies $\omega$
larger than a few tens of eV/$\hbar$ give already a negligible
contribution to the integral on the r.h.s. of Eq. (\ref{dispbis}).
On the low-frequency side, however, we have a real problem
originating from  the $1/\omega$ singularity displayed by the
imaginary part of the permittivity of ohmic conductors. As a
result $\epsilon''(\omega)$ becomes extremely large at low
frequencies, in such a way that the integral in Eq.
(\ref{dispbis}) receives a very large contribution from low
frequencies for which no optical data are available.    Since
truncation of the integral to the frequency $\omega_{\rm min}$
results in a large error,
one is forced to extrapolate the dielectric function
$\epsilon''(\omega)$ to frequencies $\omega < \omega_{\rm min}$,
where optical data are not available, to evaluate the integral for
$\omega<\omega_{\rm min}$.  As a rule, the extrapolation is done
using the simple Drude model \be \epsilon_{\rm
Dr}(\omega)=1-\frac{\omega_p^2}{\omega(\omega+ i
\gamma)}\;,\label{drude}\ee where $\omega_p$ is the plasma
frequency, and $\gamma$ is the relaxation frequency. Typical
values for these parameters for gold are $\omega_p=9$ eV/$\hbar$
and $\gamma= 35$ meV/$\hbar$ \cite{lambrecht}. As it is well
known, the Drude model is expected to provide a  reasonable
approximation to the permittivity of ohmic conductors for low
frequencies, and the hope is that the error  resulting from  use
of the Drude extrapolation is not too large. According to this
procedure, the quantity $\epsilon(i \xi)$ in Eq. (\ref{dispbis})
is computed as \be \epsilon(i \xi)=1+\epsilon_{\rm cut}(i
\xi)+\epsilon_{\rm expt}(i \xi)\;,\label{spliteps}\ee where
$\epsilon_{\rm cut}(i \xi)$ is calculated using the Drude
extrapolation for $\epsilon(\omega)$ in the unaccessible frequency
range $\omega < \omega_{\rm min}$, while $\epsilon_{\rm expt}(i
\xi)$ is calculated using the experimental optical data
$\epsilon_{\rm expt}''(\omega)$, according to the following
formulas: \be \epsilon_{\rm cut}(i \xi)=\frac{2}{\pi}
\int_0^{\omega_{\rm min}} d\omega \frac{\omega \,\epsilon_{\rm
Dr}''(\omega) }{\omega^2+\xi^2}\;,\ee and \be \epsilon_{\rm
expt}(i \xi)=\frac{2}{\pi} \int_{\omega_{\rm min}}^{\infty}
d\omega \frac{\omega \,\epsilon_{\rm expt}''(\omega)
}{\omega^2+\xi^2}\;\label{epsexpt}.\ee Later on, we shall denote
by a subscript ${\rm KK}$ estimates for the permittivities and
Casimir forces obtained by using  Eqs.
(\ref{spliteps}-\ref{epsexpt}).  The large weight of the Drude
term in Eq. (\ref{spliteps}) has been clearly recognized recently
\cite{sveto}.
  Using the handbook data,
it has been estimated (see Fig. 10 of \cite{sveto} and Fig.
\ref{epscut} below) that in the interval of imaginary frequencies
$\xi$ from 0.1 to 10 eV/$\hbar$ the quantity $\epsilon_{\rm cut}(i
\xi)$ gives the dominant contribution to $\epsilon(i \xi)$  for
$\xi< 4$ eV/$\hbar$, representing over 90 $\%$ of the total
magnitude of $\epsilon(i \xi)$ for $\xi=0.1$ eV/$\hbar$. We remark
that the situation does not  improve too much if the data interval
is extended to longer wavelengths. For example, the authors of Ref. \cite{sveto}
measured by ellipsometry the dielectric functions of several gold
films for wavelengths from 33 to 0.14 $\mu$m, corresponding to the
frequency interval from about 35 meV/$\hbar$ to 10 meV/$\hbar$.
Despite the fact that IR measurements were extended till photon
energies that are roughly one fourth smaller than
  the handbook data, still $\epsilon_{\rm cut}(i \xi)$
provides a large fraction of $\epsilon(i \xi)$, about 60 $\%$ for
$\xi=0.1$ eV/$\hbar$ and about 40 $\%$ for $\xi=1$ eV/$\hbar$ (see thick solid line of Fig. \ref{epscut}).

To give a further sense of the importance of the Drude
extrapolation for the Casimir effect, we have estimated the error
which is made in the Casimir pressure $P(a,T)$ between two gold
plates if one neglects altogether the contribution of the Drude
extrapolation in Eq. (\ref{spliteps}). As it is well known, the
Casimir force per unit area $P(a,T)$ between two identical
plane-parallel homogeneous and isotropic plates, placed in vacuum
at a distance $a$ is provided by the following Lifshitz formula
\cite{lifs} (a minus sign corresponds to an attraction between the
plates): \be P(a,T)=-\frac{k_B T}{ \pi} \sum_{n \ge
0}{\,'}\sum_{\alpha}\int \!\! d{ k_{\perp}} { k_{\perp}} q_n\,
\left(\frac{e^{2 a q_n}}{r_{\alpha}^2({ i} \xi_n,{ k_{\perp}})} -1
\right)^{-1}
  ,\label{lifs} \ee where the prime over the
$n$-sum means that the $n=0$ term has to be taken with a weight
one half, $\alpha={\rm TE,TM}$ is the polarization,  ${
k_{\perp}}$ denotes the magnitude of the projection of the
wave-vector onto the plane of the plates and $q_n
=\sqrt{k_{\perp}^2+\xi_n^2/c^2}$, where $\xi_n= 2 \pi n\,k_B T
/\hbar$ are the Matsubara frequencies. The quantities $
r_{\alpha}({i} \xi_n,{ k_{\perp}})$ denote the familiar Fresnel
reflection coefficients of the slabs for $\alpha$-polarization,
evaluated at imaginary frequencies $i \xi_n$. They have the
following expressions: \be r_{\rm TE}({ i} \xi_n,{
k_{\perp}})=\frac{q_n-k_n}{q_n+k_n}\;,\label{refTE}\ee \be r_{\rm
TM}({i} \xi_n,{\bf k_{\perp}})=\frac{\epsilon({ i} \xi_n)\,
q_n-k_n}{\epsilon({i} \xi_n)\,q_n+k_n}\;,\label{refTM}\ee where
$k_n=\sqrt{k_{\perp}^2+\epsilon({i} \xi_n)\xi_n^2/c^2}$. Unless
explicitly stated otherwise, it is understood that all Casimir
pressures here and below are computed using the so-called Drude
prescription \cite{bordag} for ohmic conductors. In this scheme,
the $n=0$ mode for TE polarization contributes nothing, and the
$n>0$ terms are evaluated using the dielectric function
$\epsilon(i \xi)$ of the considered metal, without any
modification aimed at suppressing the effect of ohmic dissipation
for conduction electrons.
\begin{figure}
\includegraphics{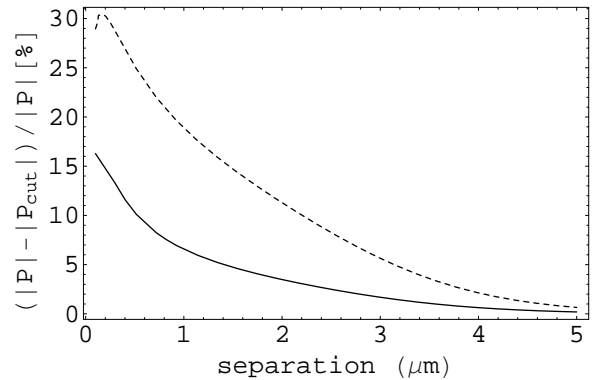}
\caption{\label{presscut}     Per cent error  in the theoretical
magnitudes of the Casimir pressure between two gold parallel
plates at room temperature, resulting from neglect of the
low-frequency contribution $\epsilon_{\rm cut}(i \xi)$ in Eq.
(\ref{spliteps}). The dashed line was computed using the handbook
data of Ref.\cite{palik} ($\omega_{\rm min}=0.125$ ${\rm
eV}/\hbar$). The solid line uses the optical data for sample 5 of
Ref. \cite{sveto} ($\omega_{\rm min}=0.042$ ${\rm eV}/\hbar$). }
\end{figure}
In Fig. \ref{presscut} we plot the fractional  difference, in per
cent, between the magnitudes $|P(a,T)|$ and $|P_{\rm cut}(a,T)|$
of the Casimir pressure between two parallel gold plates at room
temperature, that are obtained respectively if in $\epsilon(i
\xi)$ one includes or neglects the contribution of the Drude
extrapolation $\epsilon_{\rm cut}(i \xi)$ in Eq. (\ref{spliteps}).
  The dotted curve displayed in the Figure is relative
to the handbook data \cite{palik}, for which $\omega_{\rm
min}=0.125$ eV/$\hbar$. Extrapolation to low frequencies was done
using the commonly used values for the Drude parameters quoted in
\cite{lambrecht}, namely $\omega_p=9$ eV/$\hbar$ and $\gamma=35$
meV/$\hbar$. The solid curve in Fig. \ref{presscut} was instead
computed using the optical data relative to sample 5 of Ref.
\cite{sveto}, consisting of an annealed gold film deposited on a
mica substrate, with a thickness of 120 nm. We have selected this
sample, because of the five samples studied in Ref. \cite{sveto}
this one has the closest plasma frequency ($\omega_p=8.38$
eV/$\hbar$) to the commonly accepted value for gold. In
\cite{sveto} the optical data for this film were measured in the
frequency interval extending from $\omega_{\rm min}=0.042$
eV/$\hbar$ to $\omega_{\rm max}=9$ eV/$\hbar$. The Drude
parameters for this film were estimated in \cite{sveto} to have
the values $\omega_p=(8.38 \pm 0.08)$ eV/$\hbar$ and $\gamma=(37.1
\pm 1.9)$ meV/$\hbar$. When considering this film, the handbook
data from   \cite{palik} were used to extend the data of
\cite{sveto} above 9 eV/$\hbar$. We see from Fig. \ref{presscut}
that at all separations the Drude extrapolation has a smaller
weight when the data of \cite{sveto}   are used. This is a result
of the fact, noticed already, that optical data of \cite{sveto}
extend to lower frequencies than the tabulated data. We see from
the Figure that the influence of the Drude extrapolation on the
predicted pressure is very significant for both the handbook data
and for the data of \cite{sveto}, especially at submicron
separations. For a separation $a$ of 100 nm it contributes about
thirty per cent of the total pressure in the case of the handbook
data, and more than fifteen per cent for the data of Ref.
\cite{sveto}.

In the recent literature \cite{decca,geyer} it has been claimed
that relying on the handbook data of \cite{palik} and  by
following the procedure outlined above, the magnitude of the
Casimir pressure between two parallel gold plates can be predicted
theoretically with an uncertainty of 0.5 $\%$  in the separation
range from 150 nm to 400 nm. This claim does not seem warranted to
us, because of the large weight of the Drude extrapolation in the
computation of the Casimir force. Considering the magnitude of
this contribution, a precision of half per cent at   separations
around 200 nm  can be claimed only if one is confident that the
Drude extrapolation can be used to estimate the contribution to
the Casimir pressure of unaccessible low frequencies $\omega<
\omega_{\rm min}$ with a fractional precision better than one or
two per cent, a non trivial statement indeed.   In our opinion,
the reliability of Casimir computations would greatly increase  if
we could find means of substantially reducing the weight of the
Drude extrapolation. We discussed earlier that one cannot expect
to achieve this goal by further extending to lower frequencies the
data interval. We shall demonstrate below that this goal can be
achieved instead by using alternative forms of dispersion
relations, which  strongly suppress the weight of the Drude
extrapolation in determining the dielectric function $\epsilon(i
\xi)$.

\section{Weighted Kramers-Kronig relations}

In the work \cite{bimonte} we  demonstrated that in principle the
influence  of the Drude extrapolation can be much decreased by
considering alternative dispersion relations to Eq.
(\ref{dispbis}) involving appropriate weight functions, which are
aimed at  suppressing the contribution of low frequencies
$\omega<\omega_{\rm min}$, for which optical data are not
available.

Let us see briefly how the more general dispersion relations  come
about.  One starts from the following relation, generalizing Eq.
(\ref{dispbis}), that can be easily  proved by the contour
integration method: \be \epsilon({i} \xi)=1+ \frac{2}{\pi \, f({
i} \xi)} \int_0^{\infty} d\omega \frac{\omega \,
}{\omega^2+\xi^2}{\rm
Im}[f(\omega)(\epsilon(\omega)-1)]\;.\label{generdisp}\ee The
above relation holds for any weight function $f(z)$, later
referred to as "window" function,  that is analytic in the upper
complex plane, and which satisfies there the following symmetry
property \be f(-z^*)=f^*(z)\;,\label{sym} \ee In addition, the
function $f(z)$ and the permittivity $\epsilon(z)$ should be such
that the quantity $u(z)=f(z)(\epsilon(z)-1)$ has at most a simple
pole in the origin, and vanishes at infinity faster than
$1/z^{\alpha}$ for some $\alpha>0$. The standard Kramers-Kroning
relation Eq. (\ref{dispbis}) is a special case of Eq.
(\ref{generdisp}), corresponding to the choice $f(z) \equiv 1$.
The distinctive feature of the  general dispersion
formula in Eq. (\ref{generdisp}), as contrasted to the standard KK expression Eq.
(\ref{dispbis}), is that in general  it involves both the real
and the imaginary parts of the permittivity. This is not a
problem, in principle, because both quantities can be measured,
using for example ellipsometry.  The key remark made in
\cite{bimonte} was that by using window functions that vanish
sufficiently fast in the origin and at infinity,  it is in
principle possible to strongly suppress the contribution to the
integral on the r.h.s. of Eq (\ref{generdisp}) of frequencies
outside the interval $[\omega_{\rm min},\omega_{\rm max}]$ for
which no optical data are available, in such a way that an
accurate estimate of $\epsilon(i \xi)$ can be obtained by simply
truncating the integral on the r.h.s. of Eq. (\ref{generdisp}) to
the interval $[\omega_{\rm min},\omega_{\rm max}]$. We observe
that    by changing the form of the window function $f(z)$ in Eq. (\ref{generdisp}), we
have a means of changing at will the weight of the various
spectral regions, and we can modify as well the relative weights
of the real and imaginary parts of the dielectric permittivity. It
is interesting to observe also that the windowed relations offer a
possibility of checking the quality of the optical data, and in
particular their degree of KK consistency, by verifying whether
the obtained values of $\epsilon(i \xi)$ do not change when the
window function is changed.

\subsection{Old form of the window functions}

As an example, in \cite{bimonte} we considered "window" functions
$f(\omega)$ of the following  form \be f(z)= z^{2 p+1}\left[
\frac{1}{(z-w)^{2 q+1}} +\frac{1}{(z+w^*)^{2 q+1}}
\right]\;,\label{winfun}\ee where $w$ is an arbitrary complex
frequency such that ${\rm Im}(w) <0$, and $p$ and $q$ are
non-negative integers such that $0 \le p \le q$. A practical
application of the above window functions was made in Ref.
\cite{geyer}, where they were used to compute the permittivity
$\epsilon(i \xi)$ of gold, on the basis of the tabulated handbook
data  \cite{palik}. The used windowed parameters were the same
that had been used earlier in \cite{bimonte} i.e. $w=(1-2 i)$
eV/$\hbar$, $p=1$ and $q=3$. The obtained results were very
unsatisfying, since large deviations from the KK results were
found and, worse than this, unacceptable negative values for
$\epsilon(i \xi)$ were found for values of $\xi$ in the interval
from 2.44 to 2.92 eV/$\hbar$, and again for $\xi >$ 7.8
eV/$\hbar$.  The authors of \cite{geyer}  commented that the
problem was presumably determined by an exceeding propagation of
experimental errors affecting the optical data by the windowed
relations.

\begin{figure}
\includegraphics{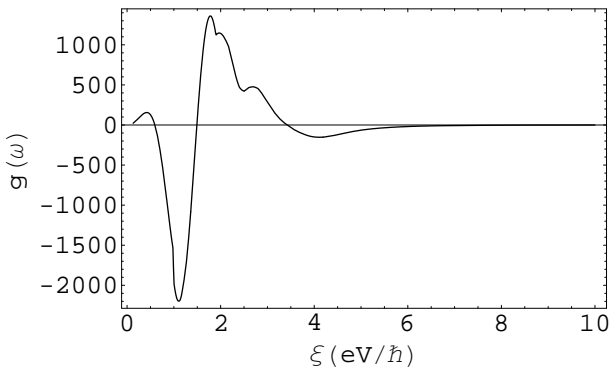}
\caption{\label{gquafig}   Plot of the quantity $g(\omega)$
relative to the handbook optical data for gold of Ref.
\cite{palik}, for $\xi=2.5$ eV/$\hbar$. The window function $f(z)$
is as in Eq. (\ref{winfun}), for window parameters $p=1$, $q=3$
and $w=(1-2 i)\,{\rm eV}/\hbar$. }
\end{figure}

After closer inspection we realized that the problem  originated
from the inconvenient choice of the analytic form of the window
function Eq. (\ref{winfun}) that we made in our first work. While
a more systematic investigation of the problem is postponed until
the next Section, we can easily understand  what is wrong with the
window functions in Eq. (\ref{winfun}) by considering the quantity
$g(\omega)$ \be g(\omega)=\frac{2}{\pi\, f({ i} \xi)} \frac{\omega
\, }{\omega^2+\xi^2}{\rm Im}[f(\omega)({
\epsilon}(\omega)-1)]\;,\label{gfun}\ee  whose integral along the
positive real axis should reproduce, according to Eq.
(\ref{generdisp}), the permittivity $\epsilon(i \xi)$-1.  In Fig.
\ref{gquafig} we plot the function $g(\omega)$, using for
$\epsilon(\omega)$ the handbook data. The displayed curve
corresponds to $\xi=2.5$ eV/$\hbar$, and to the window parameters
$p=1$, $q=3$ and $w=(1-2 i)\,{\rm eV}/\hbar$. The chosen value of
$\xi$ is one of those for which negative values of $\epsilon(i
\xi)$ were found in \cite{geyer}.
We see from the Figure that   $g(\omega)$ attains large positive
and negative values. This implies that even slight errors in the
optical data  may strongly  affect the delicate balance between
positive and negative regions in the integral on the r.h.s. of Eq.
(\ref{generdisp}), leading to large errors in the obtained value
of $\epsilon(i \xi)$. That the estimates of $\epsilon(i \xi)$ are
indeed very sensitive to errors in the optical data will be
demonstrated more thoroughly in the next Section. Here, we remark
that this instability can partially cured by using different
values for the window parameter $w$, such that the quantity
$g(\omega)$ displays lower peaks.
However, we prefer  to look for a more radical solution of the
problem, by considering alternative forms of the window functions,
which are presented in the next Section.

\subsection{Improved choice of the window function}

A highly desirable feature displayed by the standard KK relation
Eq. (\ref{dispbis}) is that the integrand on its r.h.s. is a
positive definite quantity, since the imaginary part
$\epsilon''(\omega)$ of the permittivity of all materials is
positive. This property of the integrand is  essential in ensuring
robustness of the integral with respect to small errors in the
optical data, which permits to estimate reliably the permittivity
$\epsilon(i \xi)$ along the imaginary frequency axis, provided
only that  data are available in a sufficiently large frequency
interval. This led us to wonder whether there exist choices of the
window function $f(z)$ such that the integrand on the r.h.s. of
the generalized relations Eq. (\ref{generdisp}) has a definite
sign, in such a way that the estimate of $\epsilon(i \xi)$ becomes
more robust, while at the same time preserving the key feature of
the generalized dispersion relations of suppressing the
contribution of low frequencies, for which no optical data are
available. Achieving this goal is not so easy  because differently
from the standard KK relation, the integrand on the r.h.s. of Eq.
(\ref{generdisp}) has no definite sign in general, as it can be
seen for example  in Fig. \ref{gquafig}. This is so because for
window functions $f(z)$ different from one, the integrand in Eq.
(\ref{generdisp}) involves both the real and the imaginary parts
of the permittivity $\epsilon(\omega)$, and it is well known that
differently from $\epsilon''(\omega)$, the real part of the
electric permittivity $\epsilon'(\omega)$ has no definite sign. In
addition to this, the quest for an integrand of definite sign is
further complicated by the fact that analyticity prevents the real
and the imaginary parts of the window functions from having a
definite sign, if one insists that they should vanish both at the
origin and at infinity (see discussion in Ref.\cite{bimonte}).

\begin{figure}
\includegraphics{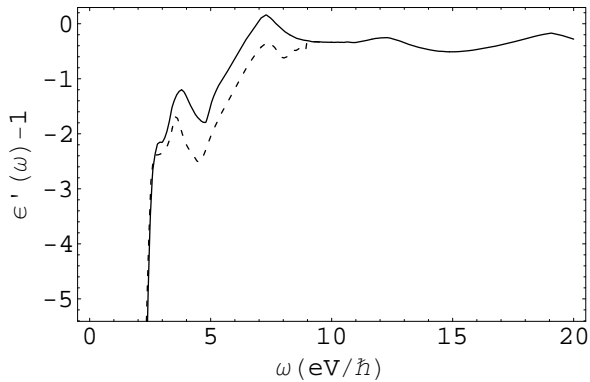}
\caption{\label{reepsminusone}   Plot of $\epsilon'(\omega)-1$ for
gold. The solid line is for the handbook data of \cite{palik}, the
dashed one for sample 5 of Ref. \cite{sveto}.}
\end{figure}

Notwithstanding these general complications, fortunately enough we
managed to find  a class of window functions for which the
integrand on the r.h.s. of Eq. (\ref{generdisp}) is in fact
positive, when ohmic conductors are considered. In order for this
to be possible,  we had  to  relax the condition that the window
functions vanish at infinity, and instead consider weight
functions that approach one at infinity. This is not a problem at
all for practical purposes, because as we said earlier optical
data extend to sufficiently high frequencies for no further
suppression to be necessary with the help of window functions on
the high frequency side. The window functions that fit our needs
have the following simple expression: \be
f(z;b)=\frac{z}{\sqrt{z^2-b^2}}\;,\label{winfunbis}\ee where $b$
is an arbitrary non-negative real frequency $b \ge 0$. As we see
these functions vanish in the origin, which ensures the desired
suppression of low frequencies in the integral on the r.h.s. of
Eq. (\ref{generdisp}). However, differently from the previous
window functions in Eq. (\ref{winfun}), these functions enjoy also
the nice feature that their real and imaginary parts are both
semi-definite functions along the positive frequency axis, since
their imaginary part is negative for $\omega < b$ and vanishes for
$\omega > b$, while their  real part is zero for $\omega < b$ and
positive for $\omega > b$.  Interestingly,  weighting functions of
the form $(\omega^2-b^2)^{-1/2}$ have been used in the past to
obtain relations connecting the refraction index $n(\omega)$ in
the interval $[0,b]$ to the extinction coefficient $k(\omega)$ in
the remainder of the spectrum (see Chapter 3 in \cite{palik}).
Insertion of Eq. (\ref{winfunbis}) into Eq. (\ref{generdisp})
results into the following formula for $\epsilon(i \xi)$: $$
\epsilon(i \xi)=
1+\frac{2}{\pi}\sqrt{1+\left(\frac{b}{\xi}\right)^2}
\left[\int_0^b d\omega
\frac{\omega^2}{\omega^2+\xi^2}\frac{1-\epsilon'(\omega)}{\sqrt{b^2-\omega^2}}\right.$$
\be +\left. \int_b^{\infty} d\omega
\frac{\omega^2}{\omega^2+\xi^2}\frac{\epsilon''(\omega)}{\sqrt{\omega^2-b^2}}\right]\;.\label{newdisp}\ee
We observe that the formula involves only the real part of the
permittivity $\epsilon'(\omega)$ for $\omega <b$, and only its
imaginary part $\epsilon''(\omega)$ for $\omega>b$. The integrand
involving $\epsilon''(\omega)$  on the r.h.s. of the above formula
is positive definite, while the integrand  involving
$\epsilon'(\omega)$ is positive whenever $\epsilon'(\omega)$ is
less than one. This is what we needed, because the real part of
the conductivity of ohmic conductors is characterized by the fact
of being less than one essentially at all frequencies, and surely
below the interband transition $\omega_{\rm inter}$, as it it can
be seen from Fig. \ref{reepsminusone}.
Therefore, if we choose $b$ to be any frequency  such that
$\epsilon'(\omega)$ is less than one for $\omega < b$  both
integrands in Eq. (\ref{newdisp}) are positive, similarly to the
standard KK relations.
 Contrasted with the standard KK relations, the new relation  Eq.
(\ref{newdisp}) receives a much smaller  contribution from low
frequencies. This is so because in the limit of vanishing
frequency the integrand in the first term of Eq. (\ref{newdisp})
vanishes like $\omega^2$, since $\epsilon'(\omega)$ stays finite
in the static limit, while the integrand of the standard KK
relation approaches a non-vanishing constant in the same limit,
due to $1/\omega$ singularity displayed by $\epsilon''(\omega)$.
Thus we may expect that the new relation   permits to estimate
$\epsilon(i \xi)$ more reliably than the standard KK relation
because, while enjoying the robustness of a positive kernel, the
weight of the Drude extrapolation is much smaller. In practical
applications of Eq. (\ref{newdisp}) we shall typically consider
frequencies $b$ that belong to the so-called non-relaxation region
$b \gg \gamma$, and are smaller than the interband transition
$\omega_{\rm inter}$. For such values of $b$, the conditions
$\epsilon'(\omega) < 0$ and $|\epsilon'(\omega)| \gg 1$ hold, and
therefore the values of $\epsilon'(\omega)$ should not be affected
by large experimental errors, contrary to what may happen at
frequencies for which $\epsilon'(\omega)$ nearly vanishes
\cite{geyer}.

For the purpose of numerical evaluation, the occurrence of an
integrable singularity for $\omega=b$ in the integrals on the
r.h.s. of Eq. (\ref{newdisp}) is quite inconvenient. We can easily
dispose of this singularity, however, by performing the change of
variables $\omega=b \sin y$ in the first integral on the r.h.s. of
Eq. (\ref{newdisp}), and $\omega=b \cosh y$ in the second one.
After doing this, Eq. (\ref{newdisp}) transforms into:
$$ \epsilon(i \xi)=
1+\frac{2}{\pi}\sqrt{1+\left(\frac{b}{\xi}\right)^2}
\left[\int_0^{\pi/2}\!\!\!\! d y \frac{\sin^2y}{\sin^2y+(\xi/b)^2}
\right.$$ \be \times [1-{\epsilon'(b \sin y)}]+\left.
\int_0^{\infty} \!\!\!dy \frac{\cosh^2 y}{\cosh^2
y+(\xi/b)^2}\,{\epsilon''(b \cosh
y)}\right]\;.\label{newdispbis}\ee

\section{Numerical computations}

We have tested the performance of the windowed dispersion relation
using the improved choice of the weight function  Eq.
(\ref{newdisp}) on some of the data sets for gold that have been
used recently for Casimir computations. First we consider the
handbook data \cite{palik}. Since they have been widely used to
interpret Casimir experiments, including in particular the
short-separation precise experiments \cite{decca}, we shall
reserve special attention to these data, and we shall spend time
to evidence some inconveniences of these data that must be handled
with some degree of caution.
In addition to the handbook data, we  used the data of the recent
work \cite{sveto}. In principle, these data present several
advantages over the handbook data. First of all, they were
obtained by using ellipsometry, which is a powerful optical
technique which permits to measure independently the real and the
imaginary parts of the dielectric function. Since our dispersion
relations explicitly involve the real part of the permittivity, it
is important to have good quality data for this quantity.  In
addition to this, the data of \cite{sveto} present  two desirable
features: on one hand, as we observed earlier, the data in
\cite{sveto} extend down to a frequency $\omega_{\rm min}$=38
meV/$\hbar$, which is   lower than the minimum frequency
$\omega_{\rm min}$=0.125 eV/$\hbar$ reached by the handbook data.
We shall find indeed that a lower frequency around 40 meV/$\hbar$
is    sufficient to compute with high precision the Casimir force,
using the windowed dispersion relations. The second and perhaps
more important  feature of  Ref. \cite{sveto} is that it provides
homogeneous data on well defined gold samples over the wide range
of wavelengths from 0.14 to 33 $\mu$m, which give the main
contribution to the Casimir effect. This is not so for the
handbook \cite{palik}, whose data in the important spectral region
from the near IR to the near UV combine results from two different
experiments \cite{theye,dold}, utilizing differently prepared gold
films. We shall see later on that the optical properties of these
gold films display striking differences, and therefore wrong
results may easily  result if the handbook data are used in
computations where KK consistency of the data plays an important
role.

\subsection{Suppression of the Drude contribution}

Our key motivation for introducing windowed dispersion relations
is that they strongly reduce the weight of the Drude
extrapolation, and correspondingly increase the weight of
experimental optical data, in such a way that obtained values for
$\epsilon(i \xi)$ are much less sensitive to experimental uncertainties in
the Drude parameters.  We can verify   to what extent the new
windowed relation Eq. (\ref{newdisp})  suppresses the contribution
of the Drude extrapolation by proceeding in a way similar to Ref.
\cite{sveto}. We decompose the quantity $\epsilon(i \xi)$ in Eq.
(\ref{generdisp}) in a way analogous to Eq. (\ref{spliteps}): \be
\epsilon(i \xi)=1+\epsilon_{\rm cut}(i \xi;b)+\epsilon_{\rm
expt}(i \xi;b)\;,\label{splitepsgen}\ee where $\epsilon_{\rm
cut}(i \xi;b)$ is calculated using the Drude extrapolation for
$\epsilon(\omega)$ in the unaccessible frequency range $\omega <
\omega_{\rm min}$, while $\epsilon_{\rm expt}(i \xi;b)$ is
calculated using the experimental optical data, according to the
following formulas: $$ \epsilon_{\rm cut}(i \xi;b)=$$
\be\frac{2}{\pi \, f({ i} \xi;b)} \int_0^{\omega_{\rm min}}
d\omega \frac{\omega \, }{\omega^2+\xi^2}{\rm
Im}[f(\omega;b)(\epsilon_{\rm Dr}(\omega)-1)]\;,\ee and $$
\epsilon_{\rm expt}(i \xi;b)=$$ \be\frac{2}{\pi \, f({ i} \xi)}
\int_{\omega_{\rm min}}^{\infty} d\omega \frac{\omega \,
}{\omega^2+\xi^2}{\rm Im}[f(\omega;b)(\epsilon_{\rm
expt}(\omega)-1)]\;,\label{splitexp}\ee where $f(z;b)$ is as in
Eq. (\ref{winfunbis}).
\begin{figure}
\includegraphics{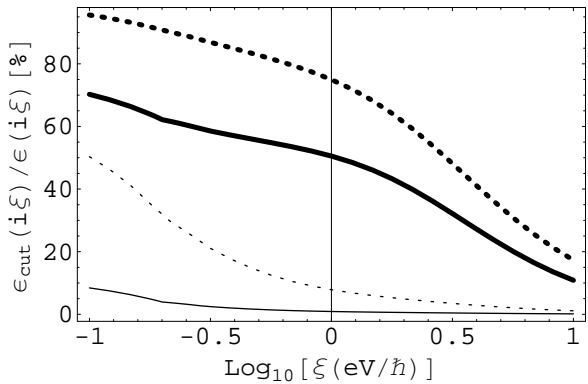}
\caption{\label{epscut} Relative contribution  of the Drude
extrapolation $\epsilon_{\rm cut}(i \xi)$  to the dielectric
function $\epsilon(i \xi)$, according to the standard KK formula
(thick lines) and to the windowed formula Eq. (\ref{newdisp})
(thin lines). Solid lines were computed using optical data of
sample 5 of Ref. \cite{sveto} (solid lines),  while the dashed
lines were computed using the handbook data \cite{palik}.}
\end{figure}
  In order to demonstrate how
well the windowed formula suppresses the Drude contribution
$\epsilon_{\rm cut}(i \xi)$, in Fig. \ref{epscut} we plot the
relative contribution of  the Drude extrapolation $\epsilon_{\rm
cut}(i \xi)/\epsilon(i \xi)$, that obtains  if one uses the
standard KK relation (thick upper solid and dashed lines) or the
generalized dispersion relation (thin lower solid and dashed
lines) with $f(z;b)$ given by Eq. (\ref{winfunbis}), for $b=1$
eV/$\hbar$. The displayed imaginary-frequency range extending from
$0.1$ eV/$\hbar$ to ten eV/$\hbar$ is the one relevant for
determining the Casimir force for plates separations larger than
100 nm. The solid lines in Fig. \ref{epscut} were computed using
the optical data relative to sample 5 of Ref. \cite{sveto}, while
the dashed lines are relative to the handbook data \cite{palik}.
We recall that in \cite{sveto} optical data were measured in the
frequency interval extending from $\omega_{\rm min}=0.042$
eV/$\hbar$ to $\omega_{\rm max}=9$ eV/$\hbar$, and so for $\omega$
larger than 9 eV/$\hbar$ we used data from Ref. \cite{palik} to
compute $\epsilon_{\rm expt}(i \xi)$. The Drude parameters used to
compute $\epsilon_{\rm cut}(i \xi)$ were as follows: for sample 5
of \cite{sveto} we used the average values quoted there, i.e.
$\omega_p=8.38$ eV/$\hbar$ and $\gamma=37.1$ meV/$\hbar$, while
for the handbook data we used the commonly used values
$\omega_p=9$ eV/$\hbar$ and $\gamma=35$ meV/$\hbar$. It is
apparent from Fig. \ref{epscut} that the generalized dispersion
relation is very effective in suppressing the contribution of the
Drude extrapolation,  in comparison  to the standard KK relation.
This is especially so for the data of \cite{sveto}, due to the
fact that they extend to smaller frequencies than the handbook
data. In this case, for the lowest displayed imaginary frequency
of 0.1 eV/$\hbar$, the relative contribution of $\epsilon_{\rm
cut}$ is reduced from   70.2  $\%$ to  8.4  $\%$, but
already for $\xi=0.5$ eV/$\hbar$ it is reduced from 55  $\%$ to
1.5  $\%$, and it becomes rapidly negligible for larger
imaginary frequencies.

We can show that thanks to  suppression of the Drude
extrapolation, experimental uncertainties in the Drude parameters
lead to much smaller uncertainties in the dielectric function
$\epsilon(i \xi)$, when  the window method is used   in comparison with the standard KK formula.
This in turn implies that the Casimir force can be predicted more
accurately by the widow approach. This is an important result
because  experimental uncertainties in the Drude parameters have a
non-negligible impact on the Casimir force, when standard KK
relations are used. To be definite, we considered the optical data
of \cite{sveto}.  In this work the Drude parameters for the used
gold films were determined by several methods, and precisely by
fitting with the Drude formula both $\epsilon(\omega)$ and the
complex refraction index $n(\omega)=\sqrt{\epsilon(\omega)}$, and
also by performing a KK analysis of the data. Depending on the
used method, the obtained values of $\omega_p$ and $\gamma$ for
the five available gold samples showed non-negligible variations,
ranging from a minimum of 1 $\%$ to a maximum of 2 $\%$ for
$\omega_p$, and from 5 $\%$ to 14 $\%$ for $\gamma$. It was found
that by itself this uncertainty in the Drude parameters entails an
uncertainty in the Casimir force  that can reach 1 $\%$ at
separations around 100 nm.
\begin{figure}
\includegraphics{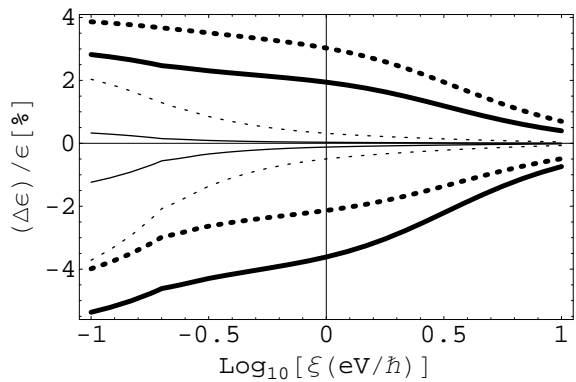}
\caption{\label{varDrude} Relative  per cent variation of the
dielectric function $\epsilon (i \xi)$ determined by small changes
in the Drude parameters, according to the standard KK formula
(thick lines) and to the windowed formula Eq. (\ref{newdisp})
(thin lines). The upper four lines correspond to a 2 $\%$ increase
in $\omega_p$, while the four lower lines correspond to a 14 $\%$
increase of $\gamma$. Solid lines were computed using optical data
of sample 4 of Ref. \cite{sveto} (solid lines), while the dashed
lines were computed using the handbook data \cite{palik}.}
\end{figure}
In order to demonstrate the  effectiveness of the window method in
reducing this source of uncertainty,  in Fig. \ref{varDrude} we
plot the relative per cent variation of the dielectric function
$\epsilon (i \xi)$ determined by  small changes in the Drude
parameters, according to the standard KK formula (thick lines) and
to the windowed formula Eq. (\ref{newdisp}) (thin lines). The
upper four lines correspond to a 2 $\%$ increase in $\omega_p$,
while the four lower lines correspond to a 14 $\%$ increase of
$\gamma$. Solid lines were computed using optical data for sample
4 of Ref. \cite{sveto} (solid lines), while the dashed lines were
computed using the handbook data \cite{palik}. In Fig.
\ref{varpres} we display the corresponding per cent variations in
the Casimir pressure between two gold plates in vacuum, at room
temperature. The lines in Fig. \ref{varpres} have the same
meanings an in Fig. \ref{varDrude}, apart from the fact that now
the 4 lower curves correspond to a 2 $\%$ increase in $\omega_p$,
while the four upper curves correspond to a 14 $\%$ increase of
$\gamma$. We clearly see  that  uncertainties in the Drude
parameters have a much smaller impact  when the window approach
 is used, as compared to the traditional KK relations. It is also clear that the window
method is more effective when applied to the data of \cite{sveto}
than for the handbook data. Indeed we see that when the former
data are used, the maximum uncertainty in the Casimir pressures
decreases from about 1 $\%$ to 0.05 $\%$, while for the handbook
data the maximum uncertainty  decreases from 0.75 $\%$ to 0.14
$\%$. Similarly to Fig. \ref{epscut}, the superior performance of
the window method when the data of \cite{sveto} are used is a
consequence of the fact that  these data extend to smaller
frequencies than the handbook data.
\begin{figure}
\includegraphics{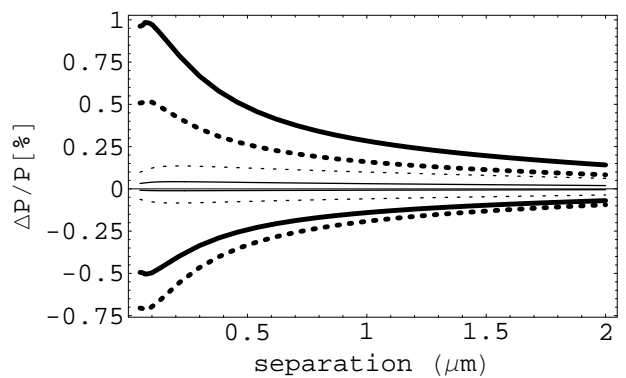}
\caption{\label{varpres} Relative  per cent variation of the
Casimir pressure between two gold plates at room temperature,
determined by small changes in the Drude parameters, according to
the standard KK formula (thick lines) and to the windowed formula
Eq. (\ref{newdisp}) (thin lines). The lower four lines correspond
to a 2 $\%$ increase in $\omega_p$, while the four upper lines
correspond to a 14 $\%$ increase of $\gamma$. Solid lines were
computed using optical data of sample 4 of Ref. \cite{sveto}
(solid lines), while the dashed lines were computed using the
handbook data \cite{palik}.}
\end{figure}


\subsection{Propagation of random errors}

The next thing that we tested is how possible errors in the
optical data are propagated by  windowed dispersion relations.
  Indeed, if we could rely on perfect optical data
with no uncertainties, the mathematical properties of analytic
functions would permit to compute the exact values of the
permittivity anywhere else in the upper complex plane, starting
from knowledge of the dielectric permittivity in any however small
interval of the real frequency axis. The existence of experimental
errors in the optical data makes this goal impossible to achieve,
and in practice  if one tries to compute the dielectric function
at points of the complex plane that are too far for the
experimental data, one inevitably ends up with estimates having
large errors. In view of this practical limitation, it is
important to investigate quantitatively how uncertainties in the
optical data are propagated by the dispersion formulae   used to
compute $\epsilon(i \xi)$. This is not an easy question to
address, because a complete answer can be given only if sufficient
information is provided on the nature the experimental errors
affecting the data.

Since neither the  handbook \cite{palik} nor Ref. \cite{sveto}
provide detailed information on the magnitude of the experimental
errors affecting their data, we performed a Monte Carlo simulation
to estimate how statistical errors in the optical data are
propagated by the windowed dispersion relations, under the
reasonable assumption of a statistical uncertainty of a few per
cent in the optical data. We proceeded as follows. Optical data
are usually presented as lists of values ${\cal S} \equiv \{n_{\rm
exp}(\omega_i),k_{\rm exp}(\omega_i)\}$, $i=1,\dots N$ for the
refraction index $n(\omega)$ and the extinction coefficient
$k(\omega)$ ($n(\omega)+i\,k(\omega)=\sqrt{\epsilon(\omega)}$), for
a discrete set of frequencies $\omega_i$: $\omega_{\rm min}
\equiv\omega_1 < \omega_2 < \dots < \omega_N \equiv \omega_{\rm
max}$. For any definite set ${\cal S}$ of data, say the handbook
data, we computed the quantity $\epsilon_{\rm expt}(i \xi)$ using
the generalized formula in Eq. (\ref{splitexp}). We then  made the
simple assumption that the experimental values of $n$ and $k$ in
${\cal S}$ had a common statistical per cent error $\delta_{\rm
exp}$. In order to estimate the corresponding   error $\delta
\epsilon_{\rm expt}(i \xi)$, we randomly generated $M$ new
hypothetical sets of data ${\cal
S}_{\alpha}=\{(n_1^{(\alpha)},k_1^{(\alpha)}),\cdots,(n_N^{(\alpha)},k_N^{(\alpha)})\}$,
$\alpha=1,2,\cdots M$.  Each data set ${\cal S}_{\alpha}$ was
generated by extracting $2N$ random numbers $\{n_i,k_i\}\;,i=1
\dots N$ from $2 N$ independent gaussian distributions, having
mean values respectively equal to the experimental values
$\{{n}_{\rm exp}(\omega_i),{k}_{\rm exp}(\omega_i)\}$, and
variances respectively equal to
 $\{({n}_{\rm exp}(\omega_i) \,\delta_{\rm exp})^2,( {k}_{\rm exp}(\omega_i)\,\delta_{\rm
exp})^2\}$.  Each data set ${\cal S}_{\alpha}$ was then used to
obtain a new estimate $\epsilon^{(\alpha)}_{\rm expt}(i \xi)$ and
the average (absolute) error ${\delta}\epsilon_{\rm expt} (i \xi)$
on $\epsilon_{\rm expt} (i \xi)$ was estimated by the formula: \be
{\delta}\epsilon_{\rm expt} (i
\xi)=\sqrt{\frac{1}{M-1}\sum_{\alpha=1}^M\left[\epsilon_{\rm
expt}^{(\alpha)}(i \xi)- {\epsilon}_{\rm expt}(i
\xi)\right]^2}\;.\ee
In our simulation we took $M=1000$. We studied by this method both
the old window functions Eq. (\ref{winfun}), and the new ones Eq.
(\ref{winfunbis}).

\begin{figure}
\includegraphics{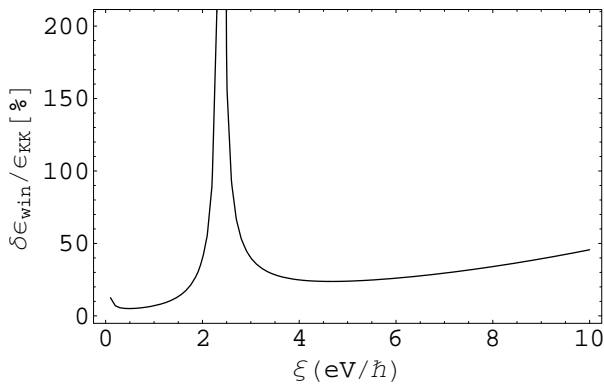}
\caption{\label{erroldwin}  Simulated uncertainty (in per cent) in
the value of the dielectric function $\epsilon(i \xi)$ versus
imaginary frequency (in eV/$\hbar$).  The dielectric permittivity
$\epsilon(i \xi)$ was computed using the handbook data
\cite{palik}, with the help of the windowed dispersion relation
for the choise of window functions in Eq. (\ref{winfun}).  The used window parameters are
 $p=1$, $q=3$ and $w=(1-2 i)\,{\rm eV}/\hbar$. In the simulation  we assumed a  3 $\%$ error
 in both $n(\omega)$ and $k(\omega)$.}
\end{figure}

We recall that by using the windowed relations with window
functions of the form given in Eq. (\ref{winfun}) the authors of
\cite{geyer} obtained unacceptable negative values for $\epsilon(i
\xi)$ in certain imaginary frequency ranges, and also for the
frequencies for which positive results were found, large
deviations from the KK values were observed. In order to explain
the findings of \cite{geyer}, we investigated first these window
functions, for the same values of the window parameters $p=1$,
$q=3$ and $w=(1-2 i)\,{\rm eV}/\hbar$ that were used in our
earlier work \cite{bimonte}, and that were subsequently considered
by Geyer et. al \cite{geyer}. Results of the Monte Carlo
simulations give support to the conjecture   made by the authors
of \cite{geyer}, that this choice of the window function
determines an  exceeding amplification of experimental errors in
the optical data, explaining the obtained negative values for
$\epsilon(i \xi)$. This can be seen from Fig. \ref{erroldwin},
where we plot the fractional random error ${\delta}\epsilon_{\rm
expt} (i \xi)/\epsilon_{\rm KK}(i \xi)$ (in per cent) for the
above choice of window parameters, assuming a  3 $\%$ error in
both $n(\omega)$ and $k(\omega)$.
As we see, the error ${\delta}\epsilon_{\rm expt} (i \xi)$ is very
large at all frequencies. The largest error occurs for values of
$\xi$ near the zero $\xi_0=2.4$ displayed by the window function
along the imaginary axis.  It is clear that estimates of
$\epsilon(i \xi)$ obtained with this choice of the window function
do not have much meaning. We  verified   that better results can
be obtained   for different values of the window parameter $w$.
For example,  for $w=-5 i$ eV/$\hbar$ the uncertainty
${\delta}\epsilon_{\rm expt} (i \xi)$ becomes less than 7 per
cent, in the entire range of $\xi$ from 0.1 and 10 eV/$\hbar$,
again assuming a  3 $\%$ error
 in both $n(\omega)$ and $k(\omega)$.

\begin{figure}
\includegraphics{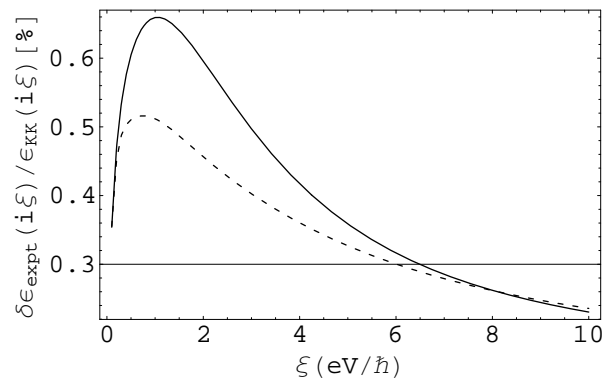}
\caption{\label{deltaeps}  Simulated uncertainty (in per cent) in
the value of the dielectric function $\epsilon(i \xi)$ versus
imaginary frequency (in eV/$\hbar$). The dielectric permittivity
$\epsilon(i \xi)$ was computed using the handbook data
\cite{palik}, with the help of the windowed dispersion relation
Eq. (\ref{newdisp}), for $b=1$ eV/$\hbar$ (solid line) and for
$b=3$ eV/$\hbar$ (dashed line). In the simulation  we assumed a 3
$\%$ error in both $n(\omega)$ and $k(\omega)$.}
\end{figure}
Even though it is possible to sensibly reduce the errors by
suitably tuning the value of $w$, in a $\xi$-dependent way, we did
not pursue further the old form of the window functions, because
much better results can be obtained by using the new window
functions given in Eq. (\ref{winfunbis}), and the associated
dispersion relation  Eq. (\ref{newdisp}). With  such a choice of
the window function, the uncertainty in $\epsilon_{\rm expt}(i
\xi)$ decrease tremendously.  This is shown in Fig.
\ref{deltaeps}, where we plot the fractional uncertainty
$\delta\epsilon_{\rm expt} (i \xi)/\epsilon_{\rm KK}(i \xi)$ (in
per cent), versus the imaginary frequency $\xi$ (in eV/$\hbar$),
for the two values of the window parameter $b=1$ eV/$\hbar$ (solid
line) and $b=3$ eV/$\hbar$ (dashed line), again assuming a  3 $\%$
error in the handbook data for $n(\omega)$ and $k(\omega)$. We see
that the uncertainty $\delta\epsilon_{\rm expt} (i
\xi)/\epsilon_{\rm KK}(i \xi)$ is now  less than 0.7 $\%$ for all
displayed imaginary frequencies. Thus our Monte Carlo simulation
gives support to the expectation that the new window functions
permit to obtain robust estimates of the dielectric function
$\epsilon(i \xi)$, thanks to the positive definite character of
the corresponding integrand on the r.h.s. of Eq.
(\ref{generdisp}). In view of these findings, we shall not
consider any longer the old window functions in the next Sections,
and we shall exclusively rely on the new ones for our next
computations.

\section{An application to tabulated data for gold}

In this Section we shall use the new window functions and the
associated  dispersion relation Eq. (\ref{newdisp}) to obtain an
estimate of dielectric permittivity $\epsilon(i \xi)$ of gold, on
the basis of the handbook data \cite{palik}.

Before we do it, an important observation on the consistency of
these data is in order. The handbook \cite{palik} quotes values
for the refraction index $n(\omega)$ and the extinction
coefficient $k(\omega)$  of gold in the frequency range from ${
\omega}_{\rm min}=0.125\, {\rm eV}/\hbar$ up to ${\omega}_{\rm
max}=9919\, {\rm eV}/\hbar$. An important feature of these data is
that they collect together  results from different experiments,
performed in different spectral regions, utilizing gold films
prepared by different procedures. In particular, data in the
infrared region of the spectrum, with frequencies in the interval
$0.125\, {\rm eV}/\hbar \le \omega \le 0.98\, {\rm eV}/\hbar$,
were taken from Ref.\cite{dold}, which used an evaporated gold
film on a polished glass substrate. Data in the frequency region
$0.6 \,{\rm eV}/\hbar \le \omega \le 6 \,{\rm eV}/\hbar$ were
taken from Ref. \cite{theye}, which used annealed films evaporated
in ultrahigh vacuum on fused silica substrates. Data in the
interval $6.199 \, {\rm eV}/\hbar \le \omega \le 26.38 \,{\rm
eV}/\hbar$ were taken from Ref. \cite{canfield}, which used
evaporated films onto polished glass substrates, in a conventional
vacuum system. Finally, data in the interval $26 \, {\rm eV}/\hbar
\le \omega \le 88 \,{\rm eV}/\hbar$ were taken from Ref.
\cite{hagemann}, which used  thin gold films evaporated onto
substrates of collodion. For the purpose of Casimir computations,
the important  spectral regions are the IR one covered by Dold's
et al. data \cite{dold}, and the region extending from the near IR
to the near UV in Theye's data \cite{theye}. At a closer
inspection, one realizes that these two sets of data are quite
inconsistent with each other, probably because of the different
deposition procedures adopted in these measurements. This  can be
appreciated by comparing the values, presented in Table 1, of the
dielectric function $\epsilon(\omega)$ computed from the two sets
of data, in the spectral region where they overlap i.e. from $0.6
\,{\rm eV}/\hbar$ to $0.9 \,{\rm eV}/\hbar$. We clearly see that
there exist large differences between the two samples.
\begin{table}
\caption{\label{tab:table1}  Values of the complex permittivity
for gold films from Ref. \cite{dold} and Ref. \cite{theye}, that
have been both included in the handbook \cite{palik}.}
\begin{ruledtabular}
\begin{tabular}{|c|c|c|}
  \hline
  $\omega$ (in ${\rm eV}/\hbar$ ) & $\epsilon(\omega )$ (Ref. \cite{dold}) & $\epsilon(\omega )$ (Ref. \cite{theye})\\
  \hline
  0.6 & -168.2 + i 23.3 & -170.7 + i 25.6 \\
  0.7 & -125.0 + i 15.6 & -165.9 + i 18.7\\
  0.8 & -96.0 + i 11.0 & -131.9 + i 12.65\\
  0.9 & -76.7 + i 7.96 & -94.3 + i 8.9\\
  \hline
\end{tabular}
\end{ruledtabular}
\end{table}

A  more striking  measure of the inconsistency between the two
data sets is obtained by estimating the respective plasma
frequencies $\omega_p$. This is an important quantity to consider,
because as we saw earlier its value has a large impact on the
magnitude of the Casimir force. We have estimated $\omega_p$ by
the following procedure. It is well known \cite{grosso} that for
frequencies $\omega$ in the non-relaxation region $\omega \gg
\gamma$ but smaller than the interband transition $\omega_{\rm
inter}$ ($\omega_{\rm inter} \simeq$ 2 eV/$\hbar$ for gold), the
real part of the electric permittivity of metals can be well
described by the formula: \be \epsilon'=1+\epsilon_{\rm
inter}-\frac{\omega_p^2}{4 \pi^2 c^2}
\lambda^2\;,\label{linfit}\ee where $\lambda$ is the wavelength
and $\epsilon_{\rm inter}$ is the contribution of interband
transitions, which for frequencies sufficiently below $\omega_{\rm
inter}$ can be considered as a constant. From the above relation,
we expect that $\epsilon'$ should vary linearly with the
$\lambda^2$, with a slope proportional to $\omega_p^2$.
\begin{figure}
\includegraphics{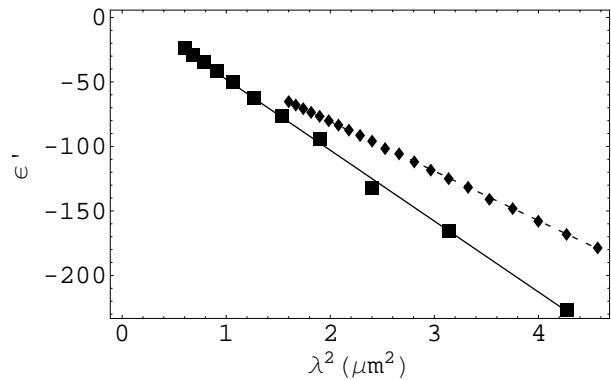}
\caption{\label{Fitsreeps}    Real part of the dielectric function
of gold in the infrared and red regions, versus square of
wavelength (with $\lambda$ in microns). Displayed data are from
Ref. \cite{theye} (squares) and \cite{dold} (diamond), which are
part of  the handbook \cite{palik}. Also displayed are the linear
fits to the data (solid and dashed lines), according to Eq.
(\ref{linfit}). The used fit parameters for the plasma frequency
are $\omega_p=9.19$ eV/$\hbar$ (solid line) and $\omega_p=7.7$
eV/$\hbar$ (dashed line line).}
\end{figure}
We determined $\omega_p$ by fitting the data in the infrared and
red regions of the spectrum. Since Theye's data \cite{theye}
extend to the interband region, we did not use $\epsilon_{\rm
inter}$ as a fit parameter for these data, but we computed its
value by means of the dispersion formula: \be \epsilon_{\rm
inter}=\frac{2}{\pi} \int_{\omega_{\rm inter}}^{\infty} \frac{d \omega}{\omega}
\, {\epsilon''(\omega)} \;,\label{epsint}\ee from which we obtained $\epsilon_{\rm inter}=5.9$. In the case Dold's et
al. data \cite{dold}, $\epsilon_{\rm inter}$ was used as a fit
parameter. We note however  that the values of $\epsilon_{\rm
inter}$ have little impact on the obtained values of the plasma
frequency $\omega_b$ because for the considered wavelengths
$|\epsilon'(\omega)| \gg \epsilon_{\rm inter}$. The fits  resulted
in the following values of the plasma frequency for the two sets
of data: $\omega_p=9.19$ eV/$\hbar$ for Theye's data and
$\omega_p=7.7$ eV/$\hbar$ for Dold's et al. data. The data and the
fitting straight lines are displayed in Fig. \ref{Fitsreeps}, from
which we see that the linear fits are very good for both sets of
data. The obtained value of $\omega_p$ for Theye's data compares
well with the value of 9 eV$/\hbar$ that was obtained on
theoretical grounds by Lambrecht and Reynaud \cite{lambrecht}, as
well as with the value of 8.9 eV/$\hbar$ that was obtained by the
authors of \cite{decca} from resistivity measurements on their
gold samples. The value of $\omega_p$ that we obtained for Dold's
et al data is consistent with the value of 7.5 eV/$\hbar$ that was
obtained by a similar method in Ref. \cite{Piro}. In the latter
work however no attempt was made to determine the plasma frequency
for Theye's data separately.
\begin{figure}
\includegraphics{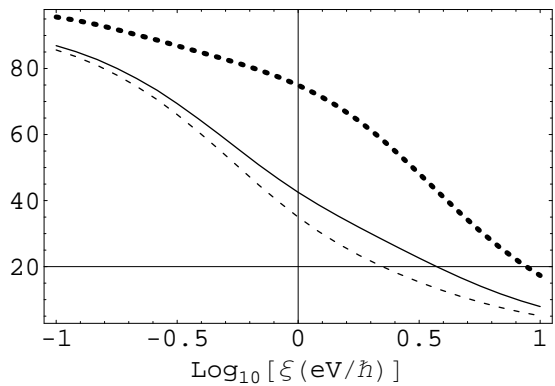}
\caption{\label{epscutTh} Relative contribution  of the Drude
extrapolation $\epsilon_{\rm cut}(i \xi)$  to the dielectric
function $\epsilon(i \xi)$ of gold, for the standard KK relation
(thick dotted line) using the complete handbook data ($\omega_{\rm
min}=0.125$ eV/$\hbar$), and the windowed dispersion formula after
exclusion of   Dold's et al. data \cite{dold} ($\omega_{\rm
min}=0.6$ eV/$\hbar$)  for $b=1$ eV/$\hbar$ (thin solid line) and
$b=1.5$ eV/$\hbar$ (thin dashed line).}
\end{figure}

Since the value of the plasma frequency for Dold's data is so much
smaller than that for Theye's data, we have decided not to use
Dold's data in our computation of the dielectric function along
the imaginary frequency axis.  The minimum frequency $\omega_{\rm
min}$ of the remaining data therefore coincides with the one
$\omega_{\rm min}=0.6$ eV/$\hbar$ for Theye's data. The Drude
parameters used to extrapolate the data below 0.6 eV/$\hbar$ were
 same as  in \cite{decca}, i.e. $\omega_{p}=8.9$ eV/$\hbar$ and
$\gamma_{p}=35.7$ meV/$\hbar$. The price  we had to pay for
excluding Dols'd data is that the contribution of the Drude
extrapolation returns to be significant, even if the windowed
dispersion relation Eq. (\ref{newdisp}) is used, as it can be seen
by comparing the thin lines in Fig. \ref{epscutTh} with the dotted
thick line in Fig. \ref{epscut}. However, thanks to the ability of
the windowed relation to suppress low frequencies, we see that
despite the  large value of $\omega_{\rm}=0.6$ eV/$\hbar$, the
weight of the Drude extrapolation in the windowed relation is
significantly less than the  one (thick dotted line in Fig.
\ref{epscutTh}) resulting from the full set of tabulated data,
beginning from 0.125 eV/$\hbar$,  when the traditional KK relation
is used.

%

\begin{figure}
\includegraphics{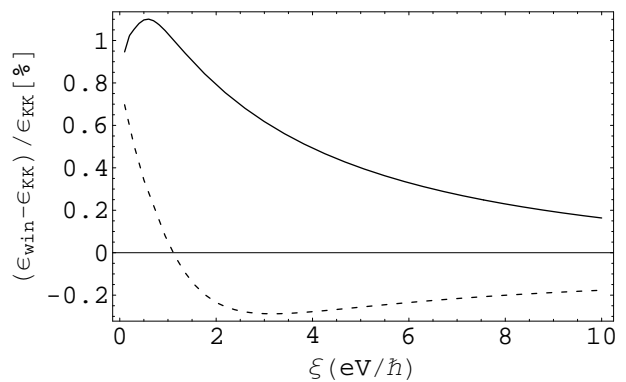}
\caption{\label{KKwindiffPalik}  Per cent difference between the
window and the KK estimates for the permittivity $\epsilon(i \xi)$
of gold, computed using the handbook data \cite{palik}. The
windowed estimate was computed using Eq. (\ref{winfunbis}), after
excluding Dold's data,   for $b=1$ eV/$\hbar$ (solid line) and
$b=1.5$ eV/$\hbar$ (dashed line). The Drude parameters used to
extrapolate the data are $\omega_p=8.9$ eV/$\hbar$ and
$\gamma=35.7$ meV/$\hbar$.}
\end{figure}

We  compared the values of  $\epsilon(i \xi)$ obtained from Eq.
(\ref{newdisp}) with those obtained from the standard KK relation
Eq. (\ref{dispbis}). We shall denote the two estimates by
$\epsilon_{\rm win}(i \xi)$ and $\epsilon_{\rm KK}(i \xi)$,
respectively.  In Fig. \ref{KKwindiffPalik} we display the
fractional difference $(\epsilon_{\rm win}(i \xi)-\epsilon_{\rm
KK}(i \xi))/\epsilon_{\rm KK}(i \xi)$ (in percent) between the two
estimates.  The solid and the dashed lines are for $b=1$
eV/$\hbar$, and for $b=1.5$ eV/$\hbar$, respectively. We remark
that the standard KK values of $\epsilon_{\rm KK}(i \xi)$ in Fig.
\ref{KKwindiffPalik} have been computed using the full handbook's
data, and therefore they coincide with the values used in
\cite{decca} for comparison of the Drude model approach with
experiments.
We note that the windowed values of $\epsilon(i \xi)$ obtained for
the two chosen values of the window parameter $b$  differ from
each other by  less than one per cent, and both differ from the KK
values by roughly the same amount. So we see that the improved
window method produces results that are substantially consistent
with the traditional method. We think however that the window
estimates obtained here are  nevertheless more reliable than the
traditional ones, because  they rely to a less extent on the Drude
extrapolation.

\begin{figure}
\includegraphics{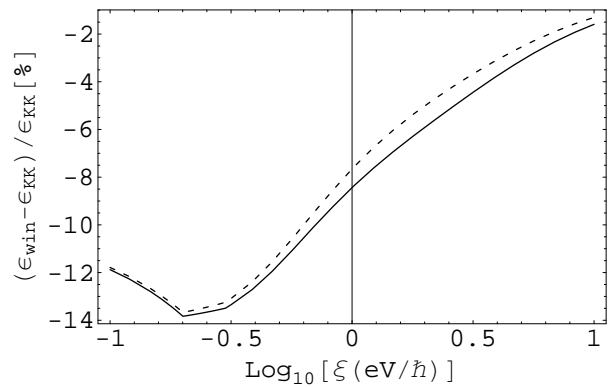}
\caption{\label{KKwindiffPalikfull}  Per cent difference between
the   window and the KK estimates for the permittivity $\epsilon(i
\xi)$ of gold, both computed using the full handbook data
\cite{palik}. The windowed estimate was computed using Eq.
(\ref{winfunbis}),
   with $b=1$ eV/$\hbar$ (solid line)
and $b=1.5$ eV/$\hbar$ (dashed line). The Drude parameters used to
extrapolate the data are $\omega_p=8.9$ eV/$\hbar$ and
$\gamma=35.7$ meV/$\hbar$.}
\end{figure}

It is interesting to see how the above results change if we use
the window method with the full set of handbook's data. This can
be seen from Fig. \ref{KKwindiffPalikfull} where we plot the
corresponding fractional difference $(\epsilon_{\rm win}(i
\xi)-\epsilon_{\rm KK}(i \xi))/\epsilon_{\rm KK}(i \xi)$ (in per
cent), the solid and the dashed lines have the same meaning as in
Fig. \ref{KKwindiffPalik}.
It is evident that after we include Dold's data the agreement
between the window and the KK estimates worsens  significantly. We
interpret this fact as a further proof of the KK inconsistency
that results from combination of Theye's and Dold's data.
\begin{figure}
\includegraphics{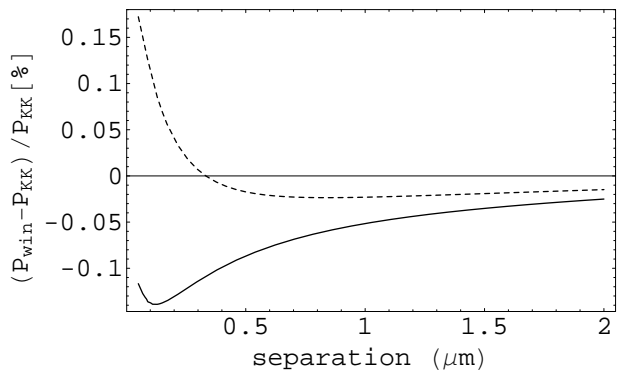}
\caption{\label{presdiff}  Per cent difference between the window
and the KK estimates for the Casimir pressure between two gold
plates, computed using the  handbook data \cite{palik}.  The
windowed pressures $P_{\rm win}$ were computed excluding Dold's
data, by using Eq. (\ref{winfunbis}) for $b=1$ eV/$\hbar$ (solid
line) and $b=1.5$ eV/$\hbar$ (dashed line). The Drude parameters
used to extrapolate the data are $\omega_p=8.9$ eV/$\hbar$ and
$\gamma=35.7$ meV/$\hbar$.}
\end{figure}

We used the windowed estimates of $\epsilon(i \xi)$ to compute the
Casimir pressure between two parallel gold plates in vacuum. We
let $P_{\rm win}(a,T)$ and $P_{\rm KK}(a,T)$, respectively, the
Casimir pressure  obtained by plugging into Eq. (\ref{lifs}) the
windowed and the traditional KK estimates for the permittivity
$\epsilon(i \xi)$. In Fig. \ref{presdiff} we plot the per cent
difference $[P_{\rm win}(a,T)-P_{\rm KK}(a,T)]/P_{\rm KK}(a,T)$
between the windowed and the KK pressures computed using the
handbook data, versus plate separation (in microns). As before,
the windowed values were computed excluding Dold's data, while the
KK pressures use the full set of data. Two different values of the
window parameter $b$ were considered, namely $b=1$ eV/$\hbar$
(solid line) and $b=1.5$ eV/$\hbar$ (dashed line). As we see, the
pressures computed by the window method differ from those obtained
by the traditional KK method by less than 0.2 $\%$ in the entire
displayed separation range. The width of the region comprised
between the solid and the dashed lines in Fig.
\ref{KKwindiffPalik} provides an indication of the uncertainty  in
the windowed prediction of the Casimir pressure, if the handbook
data are used. We estimate it to be less than 0.3 $\%$ for
separations larger than 50 nm.

\section{Application to optical data of Ref. \cite{sveto}}

We apply now the windowed dispersion relation Eq. (\ref{newdisp})
to some of the optical data for gold films of Ref. \cite{sveto}.
As we discussed at the beginning of Sec. IV these data are
particularly interesting for our purposes, because the window
method is expected to provide  very reliable results for the
dielectric permittivity $\epsilon(i \xi)$ of these samples, thanks
to the fact that in \cite{sveto} both the real and the imaginary
parts of $\epsilon(\omega)$ were measured independently
 by ellipsometry over the wide range of frequencies from $35$ meV/$\hbar$
 to about 9 eV/$\hbar$,  which includes
 longer wavelengths than the tabulated data. Of the five gold samples described
in \cite{sveto}, we  considered samples 1, 4 and 5. Samples 1 and
4 consisted of gold films deposited on a Si substrate, with
thicknesses of 400 nm and 120 nm respectively,  while sample 5
consisted of   an annealed gold film with a thickness of 120 nm
deposited on a cleaved mica substrate. The Drude parameters for
these films were determined in \cite{sveto} by several different
methods. For the convenience of the reader they are reported in
Table \ref{Drudesveto}.
\begin{table} \caption{\label{Drudesveto} Values of the Drude parameters for samples 1, 4 and 5
of Ref. \cite{sveto}.}
\begin{ruledtabular}
\begin{tabular}{|c|c|c|}
  \hline
  Sample &  $\omega_p$ [eV/$\hbar$] &  $\gamma$ [meV/$\hbar$] \\
  \hline
  1 & 6.82 $\pm$ 0.08 & 40.5 $\pm$ 2.1 \\
  4 & 8.00 $\pm$ 0.16  & 35.7 $\pm$ 5.1 \\
  5 & 8.38 $\pm$ 0.08 & 37.1 $\pm$ 1.9 \\
\hline
\end{tabular}
\end{ruledtabular}
\end{table}
We note that samples 1 and 5 (see Table II of \cite{sveto}) are
characterized, respectively, by having the smallest and largest
values of the plasma frequency $\omega_p$, and therefore the
corresponding dielectric functions $\epsilon(i \xi)$ display the
largest and the least deviation from the handbook data (see Fig.
11 of Ref. \cite{sveto}). Sample 4 on the other hand is the sample
for which the Drude parameters were found to have the maximum
uncertainties.

\begin{figure}
\includegraphics{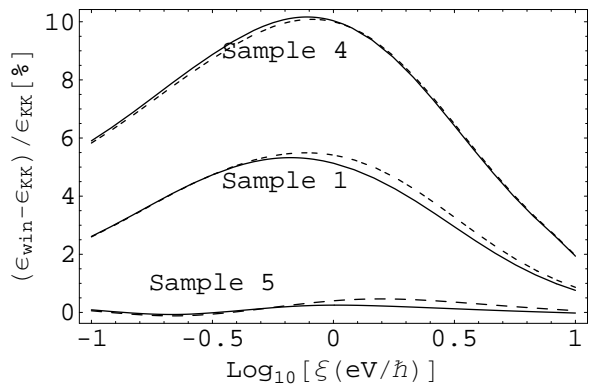}
\caption{\label{KKwindiff}  Per cent difference between the KK and
the window estimates for the permittivity $\epsilon(i \xi)$ for
samples 1, 4 and 5 of Ref. \cite{sveto}. The window function is as
in Eq. (\ref{winfunbis}). Solid lines are for $b=1$ eV/$\hbar$ and
dashed lines for $b=1.5$ eV/$\hbar$. The Drude parameters used to
extrapolate the data are listed in Table \ref{Drudesveto}.}
\end{figure}

\begin{figure}
\includegraphics{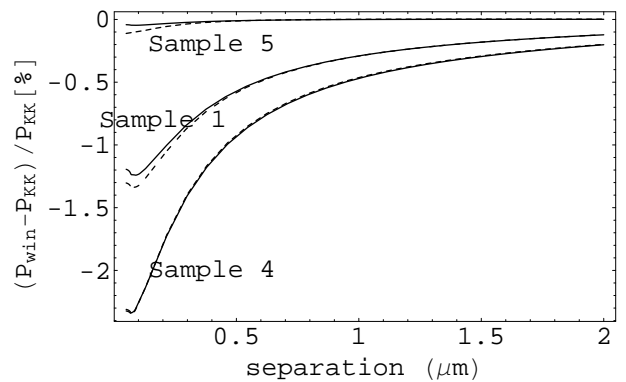}
\caption{\label{presdiffVit}  Per cent difference between the
window and the KK estimates for the Casimir pressure between two
gold plates,  for samples  1, 4 and 5 of Ref. \cite{sveto}. The
windowed pressure $P_{\rm win}$ was computed using Eq.
(\ref{winfunbis}), with $b=1$ eV/$\hbar$ (solid line) and $b=1.5$
eV/$\hbar$ (dashed line). The Drude parameters used to extrapolate
the data are   listed in Table \ref{Drudesveto}.}
\end{figure}

We evaluated the dielectric functions $\epsilon(i \xi)$ for these
three samples, by using our windowed dispersion formula, and the
traditional KK formula. In Fig. \ref{KKwindiff} we display the
fractional difference $(\epsilon_{\rm win}(i \xi)-\epsilon_{\rm
KK}(i \xi))/\epsilon_{\rm KK}(i \xi)$ (in per cent) between the
windowed and the KK estimates of the dielectric functions for
these three samples, versus the imaginary frequency. When
computing, by either method, the quantity $\epsilon_{\rm expt}(i
\xi)$ in Eq. (\ref{splitexp}), the handbook data were used for
frequencies larger than 9 eV/$\hbar$. This is not expected to have
much of an impact on the obtained results, because these large
frequencies do not contribute much to $\epsilon(i \xi)$ in the
considered range of imaginary frequencies. We observe first of all
that the window estimates for $\epsilon(i \xi)$ corresponding to
the two choices of the parameter $b$ agree within half per cent
with each other. This fact proves the good degree of KK
consistency of the optical data of \cite{sveto}.  Interestingly,
however,  the three samples display significant differences for
what concerns agreement of the window and KK estimates. While for
sample 5 the window estimates agree within less than one per cent
with the KK values, for sample 1 and sample 4 significant
deviations up to 6 $\%$ and 10 $\%$ respectively are found, the
window approach providing  systematically larger values than the
KK method. By looking at the thick solid lines in Fig.
\ref{varDrude}, one can see that the obtained discrepancies are
too large to be explained by   uncertainties in the Drude
parameters for these film, listed in Table \ref{Drudesveto}. We
speculate that the found discrepancies between the windowed and
the KK values obtained for samples  1 and 4 are related with
deviations of the  dielectric function for these films  from the
Drude model, evidenced by the presence reported in \cite{sveto} of
unexplained absorption bands in the IR, for the samples (1,2,3 and
4) deposited on Si substrates (see Fig. 7 of \cite{sveto}). The
large errors in the Drude parameters that were obtained for sample
4 probably reflect the particular inadequacy of the Drude model
for this sample. It is quite possible therefore that the KK
estimate of $\epsilon(i \xi)$, which heavily relies on the Drude
extrapolation (see Fig. \ref{epscut}), is much less accurate for
samples 1 and 4.
In Fig. \ref{presdiffVit} we display the fractional difference
$[P_{\rm win}(a,T)-P_{\rm KK}(a,T)]/P_{\rm KK}(a,T)$ (in per cent)
 between the windowed and the KK pressures versus plate separation
(in microns) for samples 1, 4 and 5. As we see the window and the
standard KK  predictions for the Casimir pressure agree within
less than 0.1 per cent in the case of sample five, while for
sample 1 and for sample 4  the window method predicts a stronger
attraction than the standard KK approach.
The theoretical uncertainty in the window prediction of the
pressure, that can be estimated by considering the difference
between the obtained pressures for the two considered values of
the window parameter $b$, is less than 0.15 per cent for all
samples and all displayed separations.

\section{Plasma prescription}

 In recent years it
has been claimed that the Drude prescription leads to results that
are in contradiction with precise measurements of the Casimir
force in the submicron region \cite{decca}. However, the recent
experiment \cite{lamor3} has been interpreted as being fully
consistent with the Drude prescription. Independently of its
accordance with experiments it has been claimed that the Drude
prescription, when used at cryogenic temperatures in the idealized
case of samples without defects, leads to a violation of the
Nernst heat theorem \cite{bezerra}. For a detailed and updated
discussion of these topics we address the reader to the monograph
\cite{bordag}. It has been argued that the above difficulties
point to a fundamental flaw in the Drude approach, connected with
relaxation processes of conduction electrons, and it has been
suggested that the above deficiencies of the Drude prescription
can be resolved if relaxation processes of conduction electrons
are neglected when evaluating Lifshitz formula.  According to this
new scheme,  the quantity $\epsilon(i \xi)$ to be  plugged in
Lifshitz formula  should not be identified with the actual
dielectric function of the metal, which includes such dissipative
processes. One should rather use  a modification of the dielectric
function, so constructed  as to disregard relaxation processes of
conduction electrons. The modified expression of the permittivity,
given in Eqs. (\ref{genpla}) and (\ref{genplabis}) below, has been
dubbed generalized plasma model, and following \cite{bordag} we
shall denote it by $\epsilon_{\rm gp}(i \xi)$. The name originates
from the fact that within this model conduction electrons are
described by the undamped plasma model $\epsilon_p(\omega)$ of
infrared physics \be
\epsilon_p(\omega)=1-\frac{\omega_p^2}{\omega^2}\;,\label{plasma}\ee
to which the Drude model Eq. (\ref{drude}) reduces when the
dissipation parameter $\gamma$ is set to zero. We note  that the
generalized plasma model is not completely immune of troubles
either, because   its $1/\omega^2$ singularity entails violation
of the Bohr-van Leeuwen theorem of classical statistical physics,
which is expected to hold for normal metals \cite{bimonte4}. In
Ref. \cite{bordag}
the following  formula  was obtained  for $\epsilon_{\rm gp}(i
\xi)$: \be \epsilon_{\rm gp}(i
\xi)=1+\frac{\omega_p^2}{\xi^2}+\frac{2}{\pi} \int_0^{\infty}
 \frac{d \omega\,\omega\, }{\omega^2+\xi^2}\, \epsilon_{\rm core}''(\omega)
  \;,\label{genpla}\ee
where the second term on the r.h.s. accounts for conduction
electrons, now described as a collisionless plasma, and
$\epsilon_{\rm core}''(\omega)$ denotes the contribution
 of core electrons.
 The plasma frequency occurring in
Eq. (\ref{genpla}) must be extracted from the data. In the case of
gold, the authors of \cite{decca} adopted the values
$\omega_p=8.9$ eV$/\hbar$ and $\gamma=0.0357$ eV$/\hbar$, that
were obtained from resistivity measurements performed at several
different temperatures on their samples.

An equivalent representation of the generalized plasma dielectric
function $\epsilon_{\rm gp}(i \xi)$, using the windowed dispersion
relation Eq. (\ref{newdisp}), can be obtained by  using the
following representation for $\epsilon_{\rm core}(i \xi)$:
$$ \epsilon_{\rm core}(i \xi)=
\frac{2}{\pi}\sqrt{1+\left(\frac{b}{\xi}\right)^2} \left[-\int_0^b
d\omega \frac{\omega^2}{\omega^2+\xi^2}\frac{\epsilon_{\rm
core}'(\omega) }{\sqrt{b^2-\omega^2}}\right.$$ \be +\left.
\int_b^{\infty} d\omega
\frac{\omega^2}{\omega^2+\xi^2}\frac{\epsilon_{\rm core}''(\omega)
}{\sqrt{\omega^2-b^2}}\right]\;. \label{altcore}\ee The equivalent
representation of $\epsilon_{\rm gp}(i \xi)$ then follows if we
replace the second term on the r.h.s. of Eq. (\ref{genpla})
representing the core contribution, by the above representation of
$\epsilon_{\rm core}(i \xi)$: $$ \epsilon_{\rm gp}(i
\xi)=1+\frac{\omega_p^2}{\xi^2}+
\frac{2}{\pi}\sqrt{1+\left(\frac{b}{\xi}\right)^2} \left[-\int_0^b
d\omega \frac{\omega^2}{\omega^2+\xi^2}\right.$$ \be \times \frac{
\epsilon_{\rm core}'(\omega)}{\sqrt{b^2-\omega^2}}+\left.
\int_b^{\infty} d\omega
\frac{\omega^2}{\omega^2+\xi^2}\frac{\epsilon_{\rm core}''(\omega)
}{\sqrt{\omega^2-b^2}}\right] \;.\label{genplabis}\ee
If the frequency $b$ is chosen to be smaller than the onset
frequency $\omega_{\rm inter}$ of interband transition, the
quantity $\epsilon'_{\rm core}(\omega)$ can be regraded as a
constant $\bar{\epsilon}_{\rm core}$. The first integral on the
r.h.s. of the above Equation can then be evaluated exactly. After
doing it we obtain the following formula for $ \epsilon_{\rm gp}(i
\xi)$:
  $$ \epsilon_{\rm gp}(i
\xi)=1+\frac{\omega_p^2}{\xi^2}+\bar{\epsilon}_{\rm core}\left[1-
\sqrt{1+\left(\frac{b}{\xi}\right)^2} \right]$$ \be +
\frac{2}{\pi}\sqrt{1+\left(\frac{b}{\xi}\right)^2}\int_b^{\infty}
d\omega \frac{\omega^2}{\omega^2+\xi^2}\frac{\epsilon_{\rm
core}''(\omega) }{\sqrt{\omega^2-b^2}}  \;.\label{genplater}\ee
The above formula can be used as a substitute for Eq.
(\ref{genpla}) to compute $\epsilon_{\rm gp}(i \xi)$ starting from
optical data.

\section{Conclusions and discussion}

In  recent years, much attention has been devoted to theoretical
and experimental studies on the thermal Casimir effect between two
metallic bodies. From the theory point of view the problem is that
of understanding whether when computing the Casimir force,
conduction electrons should be modelled as a collisionless plasma,
as it is done in infrared physics, or whether ohmic relaxation
processes should be fully  considered. The two alternatives have
been dubbed as plasma and Drude prescriptions, respectively.
Depending on the chosen approach, different predictions result for
the magnitude of the thermal Casimir force. The present
experimental status of the problem is   perplexing, because
contradictory results have been obtained by different experiments.
On one hand, a series of experiments \cite{decca} using a
micro-torsional oscillator to measure the Casimir force in the
short separation range from 160 nm to 750 nm,  were shown to be in
agreement with the plasma prescription and to rule out the Drude
prescription. Also a large distance torsion balance experiment
\cite{masuda} in which the Casimir force was measured in the
separation range from 0.48 to 6.5 $\mu$m, obtained results that
are in agreement with the assumption of ideal metal plates, and
are in contradiction with the Drude model. On the contrary a new
large distance torsional balance experiment \cite{lamor3}, probing
the Casimir force between a large spherical lens and a flat plate
in the range from 700 nm to 7 $\mu$m,  was claimed to be fully
consistent with the Drude prescription, and to rule out the plasma
model.

Motivated by the puzzling problem of the thermal Casimir effect,
in the present paper we have developed mathematical tools that
permit to obtain   reliable predictions for the Casimir force in
experiments using metallic plates, like those described above.
This is an indispensable requisite for an assessment of the plasma
versus Drude conundrum, when sub micron separations are
considered. This is so because for separations smaller than half a
micron the plasma and the Drude models predict magnitudes of the
Casimir force that differ by just a few per cent, and therefore it
is essential to make sure that the Casimir force can be actually
predicted with per cent precision or better. Our work aims at
resolving a difficulty that arises when one tries to obtain an
accurate prediction for the Casimir force, starting from optical
data for the  plates of the Casimir apparatus.
We considered specifically the case of gold films, which are used
in the experiments described above. As it is well known, in order
to compute the Casimir force, one needs to know the dielectric
function $\epsilon(i \xi)$ of the involved materials,  along the
imaginary frequency axis. This function cannot be measured
directly, and usually it is computed by means of a KK formula,
which expresses $\epsilon(i \xi)$ in terms of optical data for the
(imaginary part of the) measurable dielectric function
$\epsilon(\omega)$. The difficulty   addressed in this paper
originates from the fact that optical data   of gold are as a rule
available only for wavelengths $\lambda$ shorter than a maximum
wavelength belonging to the IR region. This is the case for
example for the widely used handbook data in \cite{palik}, where
the complex index of refraction for gold is tabulated for
wavelengths smaller than about 10 microns. The same is also true
for the recent measurements of \cite{sveto}, where the dielectric
function $\epsilon(\omega)$ for several gold films was measured by
ellipsomtery in the range from 33 to 0.14 $\mu$m. The problem  is
that knowing the dielectric function of gold for these
wavelengths, one cannot obtain an accurate estimate of the KK
integral for $\epsilon(i \xi)$. This so because the $1/\omega$
singularity displayed by the dielectric function of ohmic
conductors, makes low frequencies give a very large contribution
to the the KK integral. Until now this difficulty has been
resolved by extrapolating    available optical data towards zero
frequency by means of the simple Drude formula, which is known to
provide a reasonable approximation to the dielectric function of
ohmic conductors below the interband transition. Since the Drude
extrapolation gives a very large contribution to the KK integral,
the obtained predictions for the Casimir force are affected by a
significant uncertainty, caused by experimental errors in the
values of the Drude parameters. It was estimated in \cite{sveto}
that experimental uncertainties in the Drude parameters easily
determine a one per cent uncertainty in the Casimir force between
two gold plates at separations about 100 nm. One cannot hope to
resolve the problem by further extending optical measurements to
lower frequencies, because for this purpose one would have to
reach the Tera-Hertz region, which is practically very difficult.

Elaborating on an earlier proposal by the author \cite{bimonte},
we have shown in this paper that it is possible to resolve this
difficulty by using modified dispersion relations to evaluate
$\epsilon(i \xi)$. The new form of dispersion relations involves
certain weight functions, that we called window functions, that
are aimed at suppressing the contribution of low frequencies, for
which no optical data are available. The key feature of the new
dispersion relations, as contrasted with the standard KK formula,
is that they involve both the real and the imaginary parts of the
permittivity. This does not constitute a problem, because refined
optical techniques like ellipsometry permit to measure both
quantities accurately \cite{sveto}. The window functions
introduced in our first work produced unsatisfactory results, when
applied to the handbook data for gold \cite{geyer}. In this paper,
we demonstrated by means of a Monte Carlo simulation that the bad
performance of these window functions originated from the fact
that they lead to a large amplification of possible small
experimental errors in the optical data. In this paper we have
addressed this instability by introducing a new class of weight
functions, which differently from the old ones lead to a positive
definite kernel in the dispersion formula, in such a way that no
instability occurs anymore. We have shown that the improved choice
of window functions efficiently suppresses the contribution of low
frequencies,  in such a way that it is now possible to compute
with high precision the Casimir pressure between two
  gold plates in vacuum,  using currently available
optical data, like  the handbook data \cite{palik} and the new
data of Ref. \cite{sveto}.

When we used the handbook data,  we found it necessary to exclude
from our computations all data for frequencies less than 0.6
eV/$\hbar$, whose original source was Ref. \cite{dold}. The reason
for not using these IR data is  that the film used by these
authors appears to posses a plasma frequency of 7.7 eV/$\hbar$,
which is exceedingly smaller than the value of 9.2 eV/$\hbar$ that
we obtained from Theye's data \cite{theye}, which are part of the
handbook data from the near IR to the UV. The latter value of the
plasma frequency is very close to the commonly accepted  value of
9 eV/$\hbar$ \cite{lambrecht}. After excluding the data of
\cite{dold} the window method produces less accurate results,
because the residual handbook data that ar left over start from a
rather large frequency of 0.6 eV/$\hbar$. Nevertheless,  the
Casimir pressure can still be estimated by the window method with
an uncertainty that we estimated smaller than 0.3 $\%$. When the
Drude prescription is used, the values of the Casimir pressure
obtained by the window method agree with those obtained by the
standard approach, based on KK relations and the Drude
extrapolation, within 0.2 $\%$ for separations larger than 50 nm,
and therefore they remain inconsistent with the experimental data
of \cite{decca}.

A much higher precision can be obtained when the data of
\cite{sveto} are used. Thanks to the fact that these data extend
to longer wavelengths than the tabulated data, the uncertainty in
the windowed prediction for the Casimir pressure decreases to 0.1
$\%$, for separations larger than 50 nm. This is an order of
magnitude better than the 1 $\%$ uncertainty that was estimated in
\cite{sveto} to result from uncertainty in the Drude parameters,
within the standard KK approach.  The magnitudes of the Casimir
pressure predicted by the window method were found to be in very
good agreement with predictions based on the standard KK formula
for the mica sample of \cite{sveto}. However for the two samples
deposited on Si substrates that we considered significant
discrepancies  were found, the windowed Casimir pressures
exceeding the KK  values by more than 1  $\%$ for sample 1 and by
more than 2  $\%$ for sample 4, for separations less than 200 nm.
The discrepancy for samples 1 and 4 is too large to be explained
by uncertainty in the Drude parameters, and we attribute it to the
fact that the Drude model was shown in \cite{sveto} not to
describe well the dielectric function of these films in the IR, as
evidenced by the presence in their optical data of an unexplained
IR absorption band. It is quite possible that for this reason
predictions for the Casimir force based on the Drude model are
less accurate for these films.

In conclusion our results demonstrate that the window approach can
be used to predict the Casimir pressure between two metallic
plates very accurately, on the basis of  standard optical data
extending from the IR to the UV, without the need of further
extending optical measurements  to longer wavelengths than those
reached by current laboratory optical apparatus.  We hope that the
results of this work will help clarifying present difficulties
raised by the interpretation of experiments probing the thermal
Casimir effect.

{\it Acknowledgements} The author thanks the ESF Research Network
CASIMIR for financial support. Warm thanks are also due to G. L.
Klimchitskaya and V. M. Mostepanenko for their constructive
criticism, which greatly helped the author improving the
manuscript, and to the authors of Ref. \cite{sveto} for sharing
their data. The author acknowledges also interesting
correspondence with G. Palasantzas on the consequences of
incomplete Kramers-Kronig consistency of optical data.

\end{document}